\documentclass{aastex61}
\pdfoutput=1

\usepackage{graphicx}
\usepackage{amssymb}
\usepackage{amsmath}
\usepackage{courier}

\received{XX XX, XXXX}
\revised{X XX, XXXX}
\accepted{\today}
\submitjournal{PASP}

\shorttitle{Oxyometer for Exoplanet Atmospheric Characterization}
\shortauthors{Baker et al.}

\begin{document}

\title{The Oxyometer: A Novel Instrument Concept for Characterizing Exoplanet Atmospheres}

\correspondingauthor{Ashley D. Baker}
\email{ashbaker@sas.upenn.edu}

\author[0000-0002-0786-7307]{Ashley D. Baker}
\affiliation{University of Pennsylvania 
Department of Physics and Astronomy,  
209 S 33rd St, 
Philadelphia, PA 19104, USA}
\author{Cullen H. Blake}
\affiliation{University of Pennsylvania 
Department of Physics and Astronomy,  
209 S 33rd St, 
Philadelphia, PA 19104, USA}
\author{Sam Halverson}
\affiliation{MIT Kavli Institute for Astrophysics and Space Research
77 Massachusetts Av., 37-241,
Cambridge, MA 02139, USA}
\affiliation{University of Pennsylvania 
Department of Physics and Astronomy,  
209 S 33rd St, 
Philadelphia, PA 19104, USA}
\affiliation{NASA Sagan Fellow}

\begin{abstract}
With TESS and ground-based surveys searching for rocky exoplanets around cooler, nearby stars, the number of Earth-sized exoplanets that are well-suited for atmospheric follow-up studies will increase significantly. For atmospheric characterization, the James Webb Space Telescope will only be able to target a small fraction of the most interesting systems, and the usefulness of ground-based observatories will remain limited by a range of effects related to Earth's atmosphere. Here, we explore a new method for ground-based exoplanet atmospheric characterization that relies on simultaneous, differential, ultra-narrow-band photometry. The instrument uses a narrow-band interference filter and an optical design that enables simultaneous observing over two 0.3 nm wide bands spaced 1 nm apart. We consider the capabilities of this instrument in the case where one band is centered on an oxygen-free continuum region while the other band overlaps the 760 nm oxygen band head in the transmission spectrum of the exoplanet, which can be accessible from Earth in systems with large negative line-of-sight velocities. We find that M9 and M4 dwarfs that meet this radial velocity requirement will be the easiest targets but must be nearby (\textless 8 pc) and will require the largest upcoming Extremely Large Telescopes. The oxyometer instrument design is simple and versatile and could be adapted to enable the study of a wide range of atmospheric species. We demonstrate this by building a prototype oxyometer and present its design and a detection of a 50 ppm simulated transit signal in the laboratory. We also present data from an on-sky test of a prototype oxyometer, demonstrating the ease of use of the compact instrument design. 
\end{abstract}



\section{Introduction}
From the results of the Kepler mission it is estimated that the Transiting Exoplanet Survey Satellite (TESS) will find $10^4$ transiting exoplanets from the primary two-year mission, 3500 of which will be Neptune size or smaller \citep{TESS,2018arXiv180711129H}. A brighter selection of host stars that are smaller in radius will provide a sample of terrestrial exoplanets that will be excellent candidates for follow-up studies. \cite{TESSoccurence} predicts that for the 200,000 pre-selected stars, TESS will find 536 planets smaller than twice the radius of Earth around M dwarf hosts, while the work of \cite{ballard18} suggests the M dwarf planet yield could be 50\% higher.

In addition to radial velocity follow-up observations to determine the masses of the TESS exoplanets, atmospheric characterization will be feasible for some of these systems. Transit spectroscopy has become a very effective method for measuring the atmospheric composition of exoplanets, allowing astronomers to probe the atmospheric chemistry, interior composition, and formation environment of a growing sample of exoplanets (e.g. \citealt{seager_transitspec,madhu2014,madhu16,transitspec,kreid_transspec}). Additionally, using transit spectroscopy for detecting biosignatures in an exoplanet atmosphere could be a direct probe of potential habitability. Of the many molecules associated with life, atmospheric molecular oxygen is commonly deemed as one of the best biosignatures because of its high overall abundance, making it more accessible for detection, and its few known abiotic sources \citep{bioSig_seager}. Abiotic production of oxygen is possible, for example through the photolysis of water, but observations of other molecules such as water and methane can constrain the redox state of the atmosphere, thereby helping to rule out an abiotic origin \citep{biosig17_review,biosig_claudi17,oxygen_biosig}.

Complex, self-consistent atmospheric retrieval codes have been developed to find the most probable description of a exoplanet atmosphere provided a transit spectrum. This description includes placing constraints on the temperature-pressure profile of a planet, atmospheric chemical composition, and cloud-content of an exoplanet atmosphere \citep{retrieval_bart,retrieval_exotransmit,retrieval_madhu09,retrieval_lee13,retrieval_benneke}. Looking forward, the field has developed many sophisticated atmospheric retrieval codes that are ready for application to transmission spectra  of TESS targets. However, the challenge will be in the ability to obtain transit spectra with sufficient signal-to-noise and fidelity. The community will heavily rely upon the James Webb Space Telescope (JWST) and ground-based extremely large telescopes (ELTs) to perform observations of the most attractive TESS targets for atmospheric follow-up. Provided the many targets TESS will bring and the substantial observing time that some systems will require, JWST will only be able to target a small fraction of the most promising systems that will be selected based primarily on the results of ground-based follow-up studies \citep{morley17,barstow2015a,kempton18,2016ApJ...817...17G}. Some of these targets may be too bright for JWST and spectral information reaching into the optical wavelengths will be limited.

So far, the field has relied on pre-existing instruments operating in novel observing modes, combined with sophisticated analysis pipelines, to produce high precision (\textless 100 ppm) transmission spectra. This is particularly true of the Hubble Space Telescope (HST), which has been a key player thanks to its high-resolution spectra that are free of telluric lines and other atmospheric effects that ground based telescopes must carefully calibrate out \citep{charb_sodium_hst,hubble_deming_ATMreview,hubble_kreid_ATMreview,hst_precision,hubble_hd189}. Nonetheless, several ground based instruments have also successfully detected various species in hot-Jupiter atmospheres. In particular, sodium has been repeatably measured with several authors showing that they can successfully calibrate out Earth's sodium lines \citep{groundNA_redfield08,groundNA_khala15, groundNA_snellen08}. 
While well-calibrated spectrographs are the most competitive instruments for performing transit spectroscopy due to their high spectral resolution and large wavelength grasp, typical spectrographs suffer from low throughput and additionally require careful temperature and pressure stabilization to ensure a reliable wavelength calibration. In pushing towards the detection of molecular species in terrestrial-size exoplanet atmospheres, high instrumental throughput is crucial since operation at the photon-noise limit is key for extremely precise photometric measurements. 

Filters with very narrow bandpasses are also useful for making observations that capture the information content of narrow spectral features. A notable example showing the capabilities of transit spectrophotometry is the instrument OSIRIS on the Gran Telescopio Canarias (GTC) telescope on La Palma. With this instrument, \cite{osiris} demonstrated the detection of potassium in the hot-Jupiter XO-2b using a tunable filter with a 1.2 nm FWHM overlapping the potassium absorption feature in the exoplanet atmosphere. Although not performed simultaneously, the ability to tune the filter provided coverage over four narrow bands, allowing for the detection of the 766.5 nm potassium line. This shows the potential for ground-based spectrophotometry to aid atmospheric studies by focusing on the detection of one atmospheric species with carefully-placed, narrow filters. 

Fabry-Perot etalons are particularly well suited for transit spectrophotometry since they can be custom made to any UV to mid-IR wavelengths with a width as narrow as an Angstrom and peak throughput as high as 98\%. As demonstrated by \cite{osiris}, they can also be tuned to different central wavelengths by changing the cavity spacing. A second way to change the effective central wavelength of the filter is to change the incidence angle of the light \citep{tunablefilters_roche75}. In this case the bandpass is blueshifted as the angle of incidence of the incoming light is tilted away from normal incidence. Some consideration has been given to utilizing this technology for an instrument with the goal of achieving photon-limited, ultra narrow-band photometry \citep{narrowPhot_colon10,osiris,narrowPhot_Sioux}. However, no instruments have been developed that achieve the simultaneous observation of a star through two bandpasses that have \textless 1 nm full width at half maximum (FWHM).

We present the design of a new ultra narrowband photometer that is capable of performing simultaneous, differential measurements of a star in 0.33 nm bandpasses spaced 1 nm apart using an interference filter centered at 607.3~nm. We consider the capabilities of an instrument of similar design but in the case of a custom filter centered at the oxygen bandhead at 760 nm. Assuming this filter profile we explore the ability to detect oxygen in an Earth-like exoplanet assuming we can achieve simultaneous observations of a star in two bandpasses: one bandpass `on' the signal overlapping the oxygen 760 nm absorption region (present in a transmission spectrum assuming the target exoplanet has Earth-like levels of oxygen in its atmosphere) and the other bandpass `off' the signal so it is not overlapping the oxygen bandhead. We call this instrument an oxyometer and test this instrument's photometric performance in the laboratory and on sky using a commerical, off-the-shelf ultra-narrow interference filter centered at  607.3~nm.

In \S \ref{sec:design} we present the design concept of our theoretical oxyometer. We calculate the expected signal from an Earth-like exoplanet orbiting host stars with various radii in \S \ref{sec:signal}. We consider noise sources for our instrument and estimate the number of transits and observation time required to achieve a 3$\sigma$ detection of an exoplanetary oxygen signal using an oxyometer in \S \ref{sec:noise} and in \S \ref{sec:targetsize} we estimate the fraction of targets expected from upcoming surveys that would be appropriate for ground-based characterization by our instrument. In \S \ref{sec:lab} and \S \ref{sec:onsky} we describe the laboratory and on-sky setups and describe the results of a range of different photometric performance tests. In \S \ref{sec:discussion} we discuss future design directions and finally, in \S \ref{sec:conclude}, we summarize our results.


\section{Oxyometer Design Concept} \label{sec:design}
Here, we introduce the design of an instrument that, when combined with a large aperture telescope, could be able to reach the photometric precision needed to detect oxygen in the atmosphere of an Earth-like exoplanet. In formulating the design of this instrument, we seek a simple observing scheme that could perform robust and repeatable measurements of transiting planet systems from the ground. 

For this instrument, we propose using tunable Fabry-Perot filters that will allow the simultaneous observation of ultra-narrow bands on and off oxygen absorption lines in the exoplanet atmospheric spectrum. A major complication to executing photon-limited photometry is the telluric oxygen absorption in our own atmosphere at this wavelength, which has high optical depth and is variable over a range of timescales \citep{oxygen_variations}. The telluric oxygen signal is very strong with a series of discrete lines, some having large optical depths and even saturating when observing at higher airmass. This makes detection of the weak exoplanet signal very difficult for certain host star radial velocities when the Earth-bound and exoplanet oxygen features are aligned. Additionally, small variations in the strong telluric signal could easily propagate to noise in a measurement. Therefore, in order to perform ground-based characterization of an exoplanetary atmosphere by measuring a molecule such as oxygen that is also present in Earth's atmosphere, we must either precisely monitor the temporal behavior of the telluric signal or limit measurements to systems whose radial velocity separates the telluric and exoplanetary overlap in the wavelength domain. We therefore choose to only consider systems that have large radial velocities ($v_r$) directed towards Earth. 


A stellar radial velocity of -100~km/s would produce a 0.25~nm blueshift in the exoplanetary atmospheric spectrum. Exploiting this separation would be achievable with high resolution spectroscopy or extremely narrow band photometry. With such a large velocity offset, two filter bands, each with FWHM of $\sim$0.3~nm\footnote{The manufacture of a filter with a 0.3~nm FWHM is achievable by companies such as Alluxa (www.alluxa.com)}, would be sufficient to both avoid telluric oxygen absorption and still overlap with the blueshifted oxygen band in the exoplanet atmosphere, producing a detectable chromatic transit signal. By developing an optical design that allows for the simultaneous observation of the host star through a bandpass `on' the oxygen signal and a bandpass `off' the signal (separated by a nanometer in wavelength), we could achieve a design that has a larger system throughput and is simpler than a typical high resolution spectrograph (i.e. no slit illumination effects, thermal stabilization requirements, or wavelength calibrations), while maintaining similar benefits such as high wavelength resolution and the ability to perform simultaneous multi-wavelength measurements. We note that we limit ourselves to systems that would have a blueshifted signal because the blue edge of the oxygen A band is much sharper whereas on the red end the spectral lines are not as dense. A redshifted atmospheric signal would therefore only make a few weak lines accessible and would not result in a strong signal integrated over a 0.3 nm wide bandpass.

\begin{figure}
    \centering
	\includegraphics[width=0.8\columnwidth]{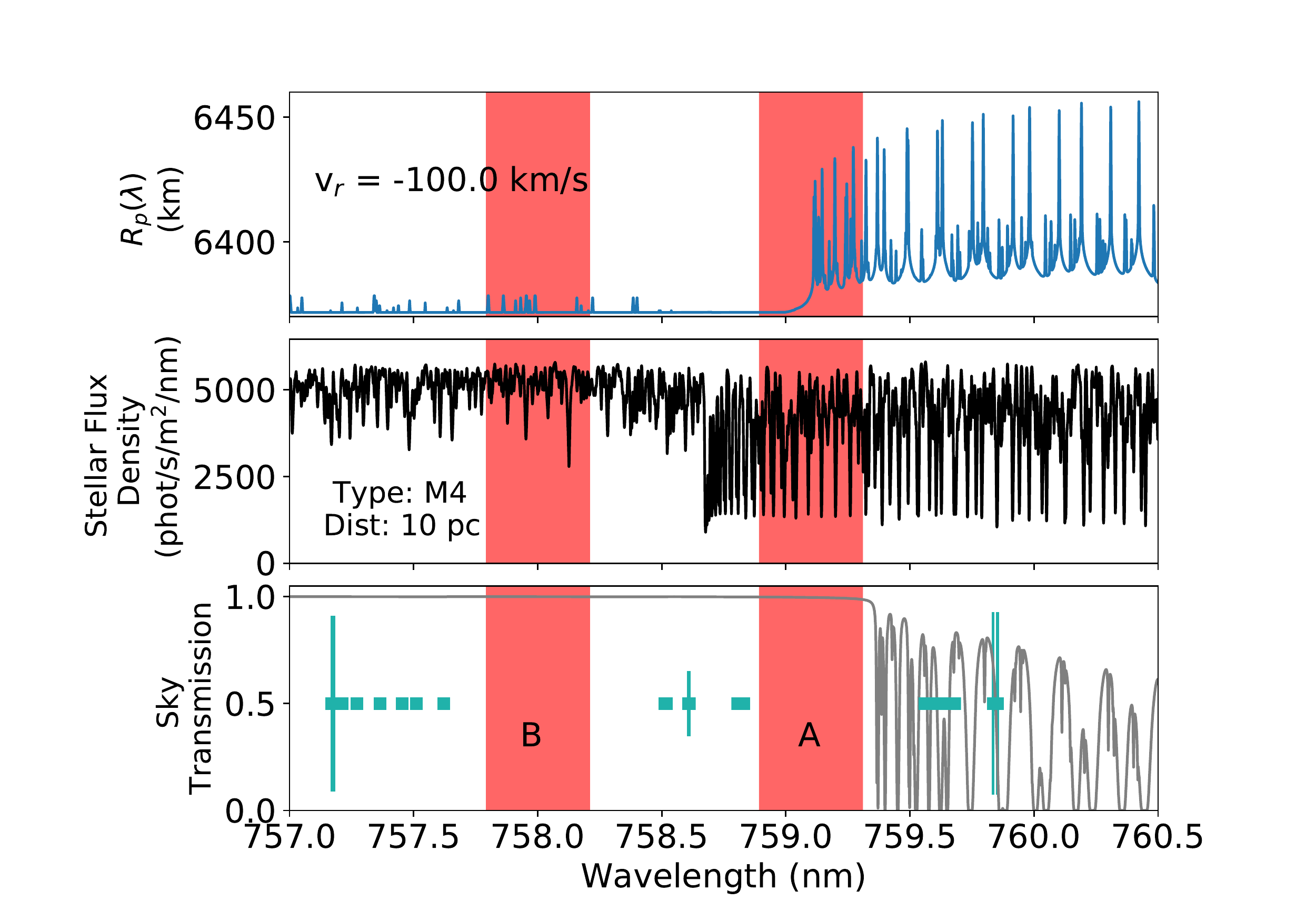}
    \caption{Exoplanet transmission spectrum plotted in blue in units of effective planet radius, $R_p$, in kilometers (top), model PHOENIX stellar spectrum in black for an M4 dwarf at a distance of 10 pc (middle), and telluric transmission spectrum (Earth's atmosphere) for an observatory at 2100 m in elevation and airmass of 1.0 shown in gray in units of transmittance (bottom). The region covered by the filter profiles used for calculations are shaded in red and labeled A and B for the on signal and off signal positions, respectively. The cyan points in the bottom panel are regions of OH emission with the x-error bars representing the FHWM of the emission while the y-error bars scale linearly with the line's central flux. The theoretical exoplanet spectrum was calculated assuming an Earth-like planet with the same radius and atmospheric contents and extent and is shown here shifted by -100~km/s, which causes the offset in the oxygen features as compared to the telluric spectrum. The stellar spectrum is similarly shifted by -100~km/s.}
    \label{fig:alluxa}
\end{figure}

In Figure \ref{fig:alluxa}, we show the proposed spectral windows of two ultra narrowband filters (shaded in red) in wavelength regimes that avoid telluric oxygen absorption and OH emission lines (plotted in cyan), which are shown in the bottom panel. We use the open source code Transit (\citealt{transit_cubillos,blecic_transit,Harrington_transit}; see \S \ref{sec:pbay}) to model the transmission spectrum of an Earth-like planet and atmosphere. We plot in the top panel of Figure \ref{fig:alluxa} this signal blueshifted by 100 km/s to mimic what the transmission spectrum would look like if the stellar system were moving towards Earth at 100 km/s. In the middle panel we show an example stellar spectrum of an M4 star also shifted by -100~km/s. Although a velocity requirement reduces the number of accessible targets, measuring the exoplanetary atmospheric signal while avoiding telluric contamination for this configuration is achievable.

The angle tuning properties of thin film filters can be utilized to avoid the need for two custom filters. With one filter centered at 759.5 nm, light from a star can be split such that half the light passes through the filter at normal incidence, while the other half passes through at an angle of incidence a few degrees off normal in order to blueshift the apparent central wavelength of the filter response. Through this method, we can achieve ultra-narrow, multiband photometry at two wavelengths that differ by around a nanometer. Using one filter instead of two reduces the cost and additionally allows us to use one CCD or other 2D detector to measure flux. This removes any potential systematic errors relating to the filters or detectors changing relative to one another in time. We discuss this optical setup more in \S \ref{sec:lab}, where we describe the prototype version that was constructed using entirely off-the-shelf components. We use the general oxyometer design described here in the following section, \S \ref{sec:signal}, to calculate the theoretical capabilities of this instrument concept.

\section{Calculations of the Oxygen Signal in an Earth-like Atmosphere} \label{sec:signal}
Observing the spectrum of an exoplanetary atmosphere in order to characterize molecular signatures can be done in several ways, exploiting various system geometries. We focus on the method of transit spectrophotometry since it is useful for targeting cool exoplanets orbiting close to their host star. In a transiting system, spectra of the host star out of transit and in transit reveal the wavelength dependent size of the planet corresponding to changing atmospheric opacity due to discrete molecular transitions. In this section we present the mathematical framework describing the transit spectral signal, which we take to be a ratio of the fluxes of the `on' and `off' bands of our theoretical instrument. We also describe how we generated an Earth transmission spectrum that we used to estimate the transit depths in each band. Using the model outputs for the atmospheric signal, we provide estimates of the signal for Earth-like exoplanets orbiting stars with various radii and system radial velocities.

\subsection{Transit Spectroscopy}
To determine the signal size of the wavelength dependent modulation of exoplanetary atmospheric absorption on transit light curve data, we assume the target star is of uniform intensity, leading to an integrated spectral flux described by:

\begin{equation}
\mathcal{F}_{out} = \frac{\pi  \mathrm{R_s}^2}{d^2} \int_0^\infty  I_{\lambda} \mathcal{T}(\lambda) d\lambda ,
\end{equation}

\noindent Here, $\mathcal{F}_{out}$ is the observed flux arriving at an observer on Earth when the planet is out of transit, $I_\lambda$ is the specific stellar intensity, $d$ is the distance between the star and Earth, $R_s$ is the radius of the star, and $\mathcal{T}(\lambda)$ is the filter profile over which the observation is made. The factor of $\pi$ is from the integration of the solid angle over a half sphere. The units of $I_\lambda$ are assumed to be in photons/m$^2$/s/nm/sr such that integrating over a wavelength gives the total flux through the bandpass, $\mathcal{T}$, in units of photons/m$^2$/s. When the planet transits the star, it will block out a fraction of the stellar flux related to the area of the planet. Assuming the effective radius of the planet, $R_p$, is a function of wavelength due to its atmosphere, we can integrate over wavelength bands to get the total flux in units of photons/m$^2$/s as:

\begin{equation}
\mathcal{F}_{in} =  \int_0^\infty \frac{\pi}{d^2} I_{\lambda} \big[ \mathrm{R_s^2} -  \mathrm{R_p^2}(\lambda) \big] \mathcal{T}(\lambda) d\lambda .
\end{equation}
Factoring out $\mathrm{R_s^2}$ yields:

\begin{equation}
\mathcal{F}_{in} =  \frac{\pi \mathrm{R_s^2}}{d^2} \int_0^\infty  I_{\lambda}  \big[ 1 -  \frac{\mathrm{R_p^2}(\lambda)}{\mathrm{R_s^2}} \big]  \mathcal{T}(\lambda) d\lambda .
\end{equation}

\noindent Here, the factor $1 -  \frac{\mathrm{R_p^2}(\lambda)}{\mathrm{R_s^2}}$ determines the modulation to $\mathcal{F}_{out,\lambda}$ due to the exoplanet atmosphere's wavelength-dependent optical depth. If we are interested in the ratio of signals taken in two narrowband filters with profiles $\mathcal{T}_A$ and $\mathcal{T}_B$ we can integrate over these two bands and take the ratio:

\begin{equation}\label{eq:flux}
\Big(\frac{\mathcal{F}_{A}}{\mathcal{F}_{B}}\Big)_{in} = \frac{\int_0^\infty I_{\lambda} \big[ 1 -  \frac{\mathrm{R_p^2}(\lambda)}{\mathrm{R_s^2}} \big]  \mathcal{T}_A(\lambda) d\lambda}{\int_0^\infty I_{\lambda} \big[ 1 -  \frac{\mathrm{R_p^2}(\lambda)}{\mathrm{R_s^2}} \big] \mathcal{T}_B(\lambda) d\lambda} .
\end{equation}

\noindent If, over the two wavelength bands, the exoplanet atmosphere has a different opacity, this ratio will deviate from the out-of-transit flux ratio. If there is no wavelength dependence of the atmosphere over the two bandpasses, the modulation factor will be the same and there will be no difference between the in transit and out of transit flux ratios. We therefore consider the signal size, $\mathcal{S}$, of our measurement to be the difference of the in and out of transit flux ratios for each of the two bands:

\begin{equation}\label{eq:signal}
\mathcal{S} = \Big(\frac{\mathcal{F}_{A}}{\mathcal{F}_{B}}\Big)_{\mathrm{out}} - \Big(\frac{\mathcal{F}_{A}}{\mathcal{F}_{B}}\Big)_{\mathrm{in}}   .
\end{equation}

We take the $A$ band to be the `on' band overlapping an exoplanet atmospheric molecular feature of interest and the $B$ band to be the `off' band overlapping a featureless portion of the exoplanetary atmospheric spectrum so that the quantity of interest, $\mathcal{S}$, is positive. 

\subsection{Transmission Spectrum Of Earth}
In order to model the transmission signal, or $R_p(\lambda)$, of an Earth-like planet and its atmosphere we use the open source code, Transit \citep{transit_cubillos,blecic_transit,Harrington_transit}. Transit\footnote{https://github.com/exosports/transit} accepts user-specified abundances and the temperature-pressure profile of an atmosphere as a function of distances above the surface of the planet. Using this information, Transit calculates the one-dimensional radiative transfer equations for the atmosphere, discretized into layers, and outputs the transmission profile as a function of the exoplanet's apparent radius. The software also handles Voigt broadening of the lines consistent with the temperature and pressure of the atmosphere. 

For Earth's abundances and temperature and pressure profile as a function of the atmospheric height above Earth's surface, we use the U.S. Standard (1976) atmospheric abundances from \cite{Earth_abundances}. We include the following species in our Earth-like atmosphere: CH$_4$, CO$_2$, O$_2$, H$_2$O, N$_2$, and N$_2$O. However, we only compute the absorption of H$_2$O and O$_2$ using the 2016 Hitran TLI line database \citep{hitran16}. We set the planet parameters of our system, including mass, radius, and surface gravity, to match those of the Earth, but vary the assumed stellar radius to consider a variety of different host stars. 

We run Transit with this atmosphere at high resolution in the wavelength range of 740 nm to 770 nm. We then convert the vacuum wavelengths generated by Transit to air wavelengths using the transformations detailed in \cite{vacuum_to_standard}. We also shift these wavelengths depending on the assumed radial velocity of the star using the nonrelativistic Doppler equation: $\lambda = \lambda_0 (1 + v_r/c)$. The output of an example calculation is plotted in the upper panel of Figure \ref{fig:alluxa} in blue. This model spectrum has been shifted by $v_r$=-100~km/s. The transit depth in regions devoid of molecular features traces the intrinsic planet radius, while the depth measured in regions containing molecular oxygen increases by a maximum of around 80 km. 

The exoplanetary transmission spectrum could also be impacted by the geometry of the system due to refraction effects limiting the ability to probe the lower atmosphere \citep{BetKal_refraction,munoz12,misra_refraction}. The effects of refraction and Rayleigh scattering are not included in the calculation of the transmission spectrum, although we expect the Rayleigh slope will be nearly flat over the considered wavelength range and we modify the spectrum afterwards to approximately account for the effects of refraction by increasing the smallest observed radii to be the planet radius plus the minimum observable atmospheric height, $H_{\mathrm{min}}$, taken from \cite{BetKal_refraction} for an Earth-like planet in its star's habitable zone. If $R_p'$ is the effective planetary radius after correcting for refraction and $R_{p,\mathrm{baseline}}$ is the radius of the planet measured to the surface, we derive $R_p'$ assuming the following conditions:

\[
    R_p'(\lambda)= 
\begin{cases}
    R_{p,\mathrm{baseline}} + H_{\mathrm{min}},     & \text{if } R_p(\lambda) < R_{p,\mathrm{baseline}} + H_{\mathrm{min}} \\
    R_p(\lambda),& \text{otherwise}
\end{cases}
\]

\noindent This approximation captures the effect of refraction to within first order since oxygen is well mixed in Earth's atmosphere, but could be improved by handling the effect in the generation of the model transmission spectrum by only integrating the opacity contributions from atmospheric layers higher than $H_{\mathrm{min}}$ above the planet surface. For an Earth-Sun analog, the bottom $\sim$13 km of the atmosphere cannot be probed in a transiting configuration due to refraction bending light rays away from a distant observer \citep{BetKal_o2}. Since we are interested in habitable planets, the smaller star-planet distances for M dwarf host stars serves to reduce the effect of refraction. In \cite{BetKal_refraction} they show that for an Earth-sized planet orbiting at a distance from its star such that it receives the same incident flux levels as present-day Earth, the atmosphere can be probed down to the surface for M5 stars and cooler. For an M0V host star, the deepest probed altitude is 5 km above the surface. The effects of refraction should be considered on a case-by-case basis for future systems that will vary in size and planet-star distance; however, for Earth-like planets in the habitable zone of a sun-like star, refraction diminishes the prospects for detecting atmospheric oxygen.

\subsection{Signal Size Calculations}\label{sec:pbay}
\begin{figure}
    \centering
	\includegraphics[width=0.7\columnwidth]{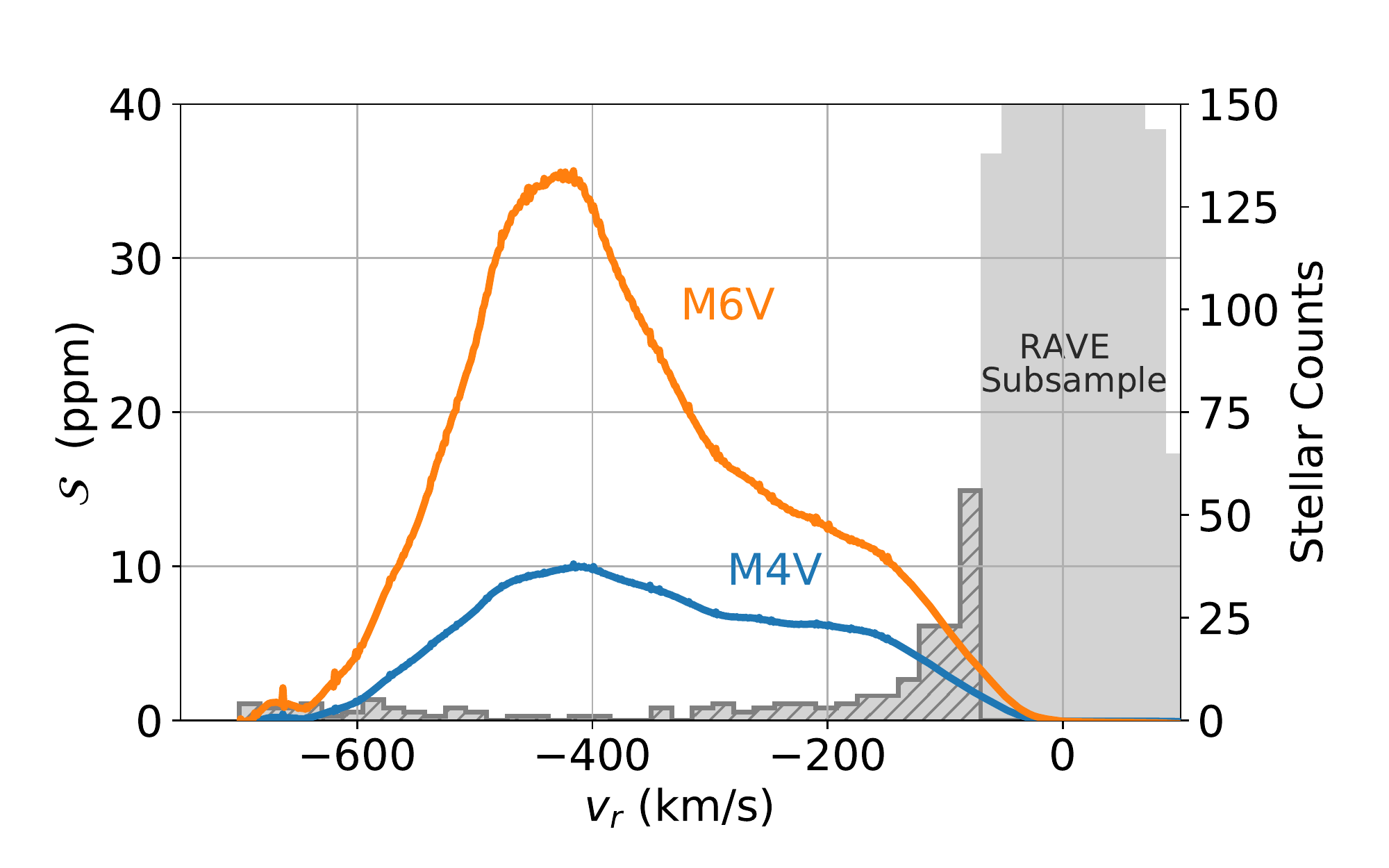}
    \caption{The signal described in \S \ref{sec:signal} versus the radial velocity of the host star for M4V (blue) and M6V (orange) stars. The oxygen signal begins to significantly overlap the filter profile when the stellar radial velocity approaches -75~km/s. After -600~km/s, the oxygen signal begins to also overlap the `off' bandpass, causing the signal to decrease again. A distribution of radial velocities from a subsample of the RAVE survey is overplotted to show that, while -75 km/s is at the tail of the distribution, there are still a significant fraction of stars that have a sufficient radial velocity for detection via this method.}
    \label{fig:f_v_vel}
\end{figure}

We use the transmission spectrum generated by Transit for $R_p(\lambda)$ corrected for refraction as described above in addition to PHOENIX stellar models \citep{phoenix} for $I_\lambda$ to calculate the expected transit depths for each of the two bands of our oxyometer for observations of an Earth-like planet  orbiting the habitable zones of stars of various radii. To do this we integrate the output Transit transmission spectrum over the chosen `on' and `off' bands (shown in Figure \ref{fig:alluxa} shaded in red) and calculate $\mathcal{S}$ according to Equation \ref{eq:flux}. In Figure \ref{fig:f_v_vel} we show the results of these calculations for M4V and M6V stars, where we have shifted the transmission spectrum to consider a range of stellar radial velocities. 

At a Doppler velocity of -100~km/s, an M6V star with an Earth-like companion will produce a differential transit signal of 5.6 ppm, while an M4V star will produce a 2.8 ppm signal. The signals increase for larger radial velocities then decreases again once the oxygen lines begin to shift into the `off' band. In Figure \ref{fig:f_v_vel} we overplot a histogram of the radial velocities of stars from a subsample of the RAVE survey \citep{rave5}. Stellar radial velocities of over 75~km/s approaching Earth are on the tail end of the distribution of observed stellar velocities. We investigate in \S \ref{sec:targetsize} what this cut on stellar radial velocities means in terms of how many transiting planets we can expect from future surveys that would be amenable to characterization with our oxyometer.

In Table \ref{tab:results} we list the radii used for determining the signals for a G2V star and M dwarfs at radial velocities of -75, -125, and -175 km/s. These are labeled as $\mathcal{S}_{75}$, $\mathcal{S}_{125}$, and $\mathcal{S}_{175}$, respectively. The transit duration assuming the planet is in the habitable zone, $\Delta T_{HZ}$, and the number of transits per year are also recorded. These values were taken from \cite{transit_durations} and \cite{rodler14}. For converting between distance and magnitude for flux measurements in \S \ref{sec:noise}, we use the absolute magnitudes from Table 5 in \cite{stellar_magnitudes}.  

We also consider performing this measurement in space where there are no constraints due to telluric oxygen. In this case the `on' photometric band can be larger and overlap the full 2 nm wide oxygen region. Widening the bandpass would also make sense for ground based observations if the target star has a large enough radial velocity towards Earth such that more than a 0.3 nm region of the oxygen band is blueshifted past telluric lines. We investigate what happens to the signal if we increase the band width to that of the full oxygen region and find that the signal remains the same to within a part per million, if comparing to a fully overlapping 0.3 nm width band. This is because the line density is roughly constant and so increasing the band averages similar proportions of spectral lines and continuum as does the smaller 0.3 nm FWHM region that we limit ourselves to for ground based observations. Although the signal stays constant, increasing the bandwidth of the filters is still beneficial since it allows for a higher flux and lower photon noise in the case of faint stars. We quantify the benefit and feasibility of performing an oxygen measurement with a 2 nm wide filter in space in \S \ref{sec:signal}.

\begin{table}
\centering
\begin{tabular}{ccccccc}
\hline
Spectral	&$R_s$ & $\Delta T_{HZ} $ &  transits  & $\mathcal{S}_{75}$  &  $\mathcal{S}_{125}$ & $\mathcal{S}_{175}$ \\ 
Type   & (R$_{\odot})$& (hrs) & per year  &  (ppm)  &    (ppm)  & (ppm) \\ \hline
G2 & 1.0 & 13.1 & 1.0 & 0.06 &  0.14  &  0.21 \\ 
M0 & 0.62 & 5.37 & 5.6 & 0.29 &  0.69  &  1.0 \\
M1 & 0.49 & 3.96 & 8.4 & 0.49 &  1.2  &  1.7 \\ 
M2 & 0.44 & 3.36 & 11.1 & 0.6 &  1.4  &  2.1 \\ 
M3 & 0.39 & 2.96 & 14 & 0.85 &  2.0  &  2.8 \\ 
M4 & 0.26 & 2.06 & 24 & 1.8 &  4.3  &  5.9 \\ 
M5 & 0.2 & 1.5 & 37   & 2.7 &  6.1  &  8.3 \\ 
M6 & 0.15 & 1.07 & 60 & 3.8 &  8.6  &  12 \\ 
M7 & 0.12 & 0.78 & 89 & 5.3 &  12  &  16 \\ 
M8 & 0.1 & 0.69 & 108 & 6.9 &  16  &  21 \\ 
M9 & 0.08 & 0.43 & 192 & 15 &  34  &  48 \\ 
\end{tabular}
\caption{Parameters assumed for G2V and M0V-M9V stars hosting Earth-like planets in their habitable zones and the resulting signal calculated for observations with an oxyometer instrument for stellar radial velocities of -75~km/s, -125~km/s, and -175~km/s. Stellar radii, transit durations, and number of transits per year for Earth-like planets in the habitable zone were taken from \cite{transit_durations}.}
\label{tab:results}
\end{table}

\subsection{Effects that Could Reduce the Signal}

\subsubsection{Clouds \& Hazes in the Atmosphere}
It is important to note that clouds and hazes in the exoplanet atmosphere can reduce the depth into the atmosphere that can be probed by transmission spectroscopy by preventing any potential modulation the lower atmosphere would have imprinted on the transmission spectrum \citep{kreidberg14,charnay15,demory13,Pont2013,knutson14}. The higher the cloud layer is in the atmosphere, the more atmosphere is obscured and cannot be probed. If the clouds are thin, or patchy, there is still a chance that a large enough portion of the atmosphere is accessible along the line-of-sight of background stellar photons for the transmission spectrum to be modulated.

Several hot Jupiter exoplanets have been characterized with transit spectroscopy and were found to have very flat atmospheric spectra \citep{sing_clouds}. It is not clear what determines the cloudiness of an exoplanet, but it should be expected that a similar fraction of terrestrial planets will have clouds \citep{cloudsMarley13}. If these planets have cloud systems similar to Earth's, then we can expect that the patchy clouds at a low atmospheric height will not block a significant portion of the atmosphere and the detection of oxygen would still be possible, particularly since oxygen is well mixed in Earth's atmosphere. However, a thick cloud layer at 10 km would, for example, still reduce the signal by completely blocking the lower atmosphere thereby reducing the modulation amplitude in the resulting transmission spectrum.

The presence of clouds obscuring the atmosphere is degenerate with the possibility of lower oxygen abundances in the exoplanet atmosphere and, with two spectral bands, an oxyometer would not have the spectral coverage to discern between the scenarios of clouds, an oxygen-free atmosphere, a high mean molecular weight atmosphere, or no atmosphere at all. Other measurements of the exoplanet at different wavelengths could help distinguish between these scenarios, particularly if other atmospheric species are observed. We perform our estimates assuming the optimistic scenario of no reduction in our signal due to the presence of clouds.


\section{Expected Photon-limited Noise Estimates} \label{sec:noise}

Here, we estimate the expected noise sources that would affect an on-sky experiment with our theoretical oxyometer design proposed in \S \ref{sec:design}. We consider instrument specific noise including photon noise, dark noise, and read noise in order to calculate estimates for various telescope and CCD specifications to determine the observational requirements to detect the signal sizes estimated in the previous section. We also discuss sources of noise external to the instrument including scintillation, other telluric emission and absorption, and noise from the host star.

\subsection{Instrument Noise Sources \& SNR Estimates}
To calculate the signal-to-noise ratio (SNR) for a measurement of a transiting Earth-like exoplanet using a theoretical oxyometer instrument, we consider photon, read, and dark noise for a range of telescope diameters. The measurement is ultimately photon noise limited and to reduce the noise to below parts per million levels while integrating over spectral bands as narrow as the 0.3 nm FWHM we are considering could result in an unrealistic integration time if the star is too faint. To determine the capabilities of our instrument we consider various stellar types and distances and determine the transit observations required for a 3$\sigma$ detection\footnote{We choose a 3$\sigma$ detection limit in order to compare to the works of \cite{rodler14} and \cite{snellen_o2}} on a range of telescope diameters. 

In the previous section we calculated our signal which is defined in Equation \ref{eq:signal}. In an observing scenario, we would measure the flux through each band split up into separate exposures, continuously measuring the flux of a star through the instrument's two ultra-narrow bands over the entire duration of the transit. We would take the ratio of the fluxes in the two bands for two time windows: when the planet is in transit and a period of similar duration when the planet is out of transit. Like before, the difference of the flux ratio for when the planet is in transit versus when it is not in transit is the ultimate measurement we aim to make. The observed signal, $\mathcal{S}$, can be rewritten as the following equation provided $i$ observations out of transit and $j$ observations in transit for a total of N and M observations each, respectively:

\begin{equation}\label{eq:obs_sig}
\mathcal{S} = \frac{1}{N}\sum_i^N\Big(\frac{F_A}{F_B}\Big)_{i,out} - \frac{1}{M}\sum_j^M\Big(\frac{F_A}{F_B}\Big)_{j,in} .
\end{equation}

Here, $F_A$ and $F_B$ are now the observed fluxes in units of photons observed in a single exposure over the $A$ and $B$ bands, which is determined by taking $\mathcal{F}$ defined in \S \ref{sec:signal} and multiplying this by the overall instrument throughput including telescope reflectivity, instrument throughput, and quantum efficiency ($\eta$), exposure time (t$_{\mathrm{exp}}$), and telescope effective area ($A_{\mathrm{tel}}$), such that $F=\mathcal{F} \mathrm{t}_{exp} \eta A_{\mathrm{tel}}$. For simplicity in these calculations, we ignore the loss of signal due to the egress and ingress but note that this slightly underestimates the required observing time to reach a 3$\sigma$ detection. For example, an Earth-like exoplanet around an M4 dwarf spends 7\% of the transit\footnote{This value was calculated using Equation 2 of \cite{2003seager_transit_duration}} in the egress and ingress, where the signal size will be a fraction of what is determined for the ''flat part" of the transit. This will be slightly alleviated since in practice the data over the whole transit can be analyzed in time bins and fit with a transit light curve function instead of combining all exposures to a single measured quantity. Structure in a time sequence of real measurements can provide insight into systematic effects that may occur that are unrelated to an atmospheric signal such as the airmass dependence in the measured flux ratio or stellar contamination due to a heterogeneous stellar disk. In the absence of systematic effects however, it is valid to sum the data across the transit window and consider the uncertainty in the measurement using the cumulative photon count. 

We use Equation \ref{eq:obs_sig} for determining the noise estimates for an observation that is realistically structured in terms of integration sequences and measured quantities. We start with the uncertainty in the flux of a single band and single exposure, $\sigma_{F}$. We assume the CCD equation and include photon noise,  background Poisson noise,  dark
noise, and read noise: 

\begin{equation}
    \sigma_F = \sqrt{ F + \mathrm{n}_\mathrm{pix}(N_B \mathrm{t}_\mathrm{exp} A_\mathrm{tel} + N_D \mathrm{t}_\mathrm{exp} + N^2_R)} \: .
    \label{eq:CCDeq}
\end{equation}

\noindent
Here, $N_B$ is the background count flux in photoelectrons, $N_D$ is the dark current, and $N_R$ is the read noise per read. We define the values used for calculating the noise in the following section. We propagate the uncertainty in each flux value, which we call $\sigma_{F_A}$ or $\sigma_{F_B}$ to denote the different filter bands, to the flux ratio for either the in or out of transit regimes. We present this for the in transit case below. 

\begin{equation}\label{eq:s_err_in}
\sigma_{in} = \Big(\frac{F_A}{F_B} \sqrt{ \Big[ \Big(\frac{\sigma_{F_A}}{F_A}\Big)^2 + \Big(\frac{\sigma_{F_B}}{F_B}\Big)^2 \Big]} \: \Big)_{in} \: .
\end{equation}

\noindent Propagating the noise in the flux ratios to the final signal gives:

\begin{equation}\label{eq:s_err}
\sigma_\mathcal{S} = \sqrt{\sigma_{out}^2 + \sigma_{in}^2}
\end{equation}

\noindent We use Equation \ref{eq:s_err} and assume that the noise is uncorrelated such that after $N=M=n_\mathrm{exp}$ exposures, $\sigma_\mathcal{S}$ is reduced by a factor of $\sqrt{n_{exp}}$ such that we can write down the signal to noise ratio as: 

\begin{equation}\label{eq:snr}
\mathrm{SNR} = \sqrt{\mathrm{n}_\mathrm{exp}} \frac{\mathcal{S}}{\sqrt{\sigma_{out}^2 +  \sigma_{in}^2}}
\end{equation}

\noindent To determine the number of exposures required to reach a certain signal to noise ratio, we set Equation \ref{eq:snr} to a value of 3 that we assume for our target signal to noise ratio and solve for n$_\mathrm{exp}$. If SNR$(\mathrm{n}_{exp} = 1)$ is the signal to noise when one exposure has been taken, then the number of exposures required to reach a SNR of 3 is

\begin{equation}\label{eq:n_exp}
\mathrm{n}_\mathrm{exp} = \Big(\frac{3}{\mathrm{SNR}(\mathrm{n}_\mathrm{exp} = 1)}\Big)^2,
\end{equation}

\noindent where SNR$(n_{exp} = 1)$ is evaluated for each stellar type and distance. Given this exposure time, t$_\mathrm{exp}$, and the duration of transit of an Earth-like planet in the habitable zone of each star, $\Delta t_{HZ}$, we can determine the corresponding number of transits, $\mathrm{n}_{\mathrm{transits}}$, needed to to reach a 3$\sigma$ detection. This is equivalent to:

\begin{equation}\label{eq:n_transit}
\mathrm{n}_{\mathrm{transits}} = \frac{\mathrm{n}_\mathrm{exp} \mathrm{t}_\mathrm{exp}}{\Delta \mathrm{t}_{HZ}}
\end{equation}

\noindent
We point out that our use of $\mathrm{n}_\mathrm{exp}$ counts a pair of exposures, one for a flux ratio taken in transit and a second taken out of transit. A factor of two on $\mathrm{n}_\mathrm{exp}$ to account for this cancels with a factor of two that would be required on $\Delta \mathrm{t}_{HZ}$ to account for each transit requiring twice the time to include the time for the out of transit observations. The total observing time however must include the factor of two on $\mathrm{n}_{exp}$. The total observing time required to reach a signal to noise ratio of three, $t_{3\sigma}$ is then equal to $2 \mathrm{n}_\mathrm{exp} \mathrm{t}_\mathrm{exp}$.

\subsection{Parameter Assumptions}
For our estimates of $\mathrm{n}_{\mathrm{transits}}$ or the total required observing time, we assume two scenarios with different instrument efficiency values: a case with $\eta=0.25$ that includes an extra factor of 50\% for an assumed beam split and an optimistic case with no beam split and with a total instrument throughput, $\eta$, of 80\%. For each case we determine the flux of each type of star using the PHOENIX models scaled to the correct I-band magnitude provided the assumed distance. To get $\mathcal{F}$ we integrate each scaled PHOENIX spectrum times $T(\lambda)$ for which we use a 0.3 nm FWHM Alluxa filter profile (or 2 nm width bandpass if specified) at the two bandpass positions A and B shown in Figure \ref{fig:alluxa}. For the calculation of the in transit fluxes, we first multiply the PHOENIX spectrum by the Earth transmission spectrum converted to percent throughput: 1-($R_p$($\lambda$)/$R_s$)$^2$. For $A_\mathrm{tel}$, we use the telescope effective areas reported in Table \ref{tab:diameters} and for t$_\mathrm{exp}$ we use 15 min. We choose a set exposure time since it would be possible to assume the use of EMCCDs that have very low read noise\footnote{Since our measurement relies upon a relative measurement between two fluxes recorded in the same exposure frame, a time varying gain across the chip as is typical in EMCCDs would not affect the measurement.} such that scanning across the CCD would avoid saturation while adding little contribution to the noise. 

For our noise parameters, $N_B$, the background count flux, is 0.18 e-/pixel/s/m$^2$, $N_D$, the dark current, is 0.22 e-/pixel/s, and $N_R$, the read noise, is taken to be 3 e-/pixel/read. We choose the background count flux based on the R band sky brightness, in magnitudes, at the Gemini Observatory for dark time\footnote{http://www.gemini.edu/sciops/telescopes-and-sites/observing-condition-constraints/optical-sky-background}, which corresponds approximately to within a week from new moon. For all telescopes we assume a focal ratio of $f$=17, which corresponds to that of the European Extremely Large Telescope (E-ELT) and compute the plate scale according to $p=206265.0/D/f$ for each telescope of diameter $D$. This gives the plate scale, $p$ in units of arcseconds/meter. We assume 18 micron square pixels, $l_\mathrm{pix}$, and use this along with the assumption of a 0.05 arcsecond star size on the detector to determine the number of pixels the star spans on the detector, $d_\mathrm{pix}$ and use this aperture diameter determine the area of pixels, n$_\mathrm{pix}$ used in Equation \ref{eq:CCDeq}. We calculate the diameter of the star in pixels according to:

\begin{equation}
  d_\mathrm{pix} = \frac{0.05 \: \mathrm{arcsec}}{p* l_{\mathrm{pix}}} 
\end{equation}

\noindent This equates to an aperture diameter of about 9 pixels for the E-ELT. We point out that the assumption of a 0.05 arcsecond star sized was guided by the diffraction limit of the ELT (0.01 arcsecond) assuming correction for atmospheric distortions and that the size of the diffraction-limited stellar image scales with telescopes aperture. However, changing the size of the star on the detector by a factor of a few to accommodate different telescope properties and seeing conditions does not significantly alter our estimates. In practice, the size of the Point Spread Functions can be optimized given the exact instrument and detector specifications to maximize stellar photons contained in the aperture while reducing read and dark noise. We do not include scintillation noise in the calculations presented here, but we address how scintillation noise affects our final SNR estimates in \S \ref{sec:scint}. 

\subsection{Sky Noise Sources}

Sources of measurement error related to Earth's atmosphere such as telluric absorption, scintillation noise, and sky emission in either of the `on' and `off' bandpasses of a differential photometer must be considered. Overlap of one of the bands with a noise source that varies on shorter timescales than that of the exoplanet transit could potentially hide the underlying exoplanetary signal. To validate that an oxyometer would be useful on ground based telescopes, we must check that the bandpass is sufficiently avoiding the telluric oxygen absorption lines, that there are no other significant absorption or emission mechanisms corresponding to other telluric species overlapping the bandpass regions, and finally that there is also no coincident sky emission. We must also evaluate the impact of differential scintillation noise across the 1~nm wavelength range of the oxyometer. We discuss each of these effects individually below.

\subsubsection{Telluric Oxygen}
Although the `on' bandpass is chosen to avoid the telluric oxygen lines, the chosen positioning still leaves slight overlap between the broad wings of the oxygen lines and edge of the filter bandpass. To check the impact that typical variations in telluric oxygen concentration would have on our signal, we simulate the effect by modifying the optical depth of oxygen and propogating the effective change in throughput that occurred in the `on' band. We find that a 10\% change in the optical depth of the telluric oxygen lines causes a 1 ppm change in the value of $\mathrm{S}$. Oxygen is well mixed in Earth's atmosphere and so the total oxygen column above the observatory is very stable. Low-level, nightly variations at this level are not very well studied. Even if we consider oxygen variations at the 1\% level over a few hours this would result in a 0.1 ppm contamination in our signal. Aside from changes due to increasing airmass that will be slowly varying and can be modeled and removed from the data, a 1\% change in oxygen concentration is not expected on few-hour timescales. We therefore can expect that changes in telluric oxygen absorption will not be a significant source of systematic uncertainty in our measurements in one night of observing. Seasonal variations of molecular oxygen have been studied and do occur up to an amplitude of 20 parts per million by volume (ppmv) \citep{oxygen_variations}. A change of 20 ppmv is a 0.01\% change in the abundance of oxygen (mixing ratio of 0.21 mol/mol). This would produce an effect \textless 0.01 ppm in our measurement. We can therefore also conclude that seasonal variations in telluric oxygen would not introduce noise in the process of co-adding transit measurements taken over long timescales. 

\subsubsection{Telluric Water}
Aside from telluric oxygen, weak telluric water vapor lines also overlap the wavelength range of interest. Several microtelluric lines that are on the level of $\sim$0.1\% in depth are present in both bandpasses. Although the precipitable water vapor at a given observing site can change by several millimeters over hour timescales \citep{camal,li2018}, these microtelluric lines are very low-level. Using a similar method to that used to check the effects of changing oxygen concentration, we find that a 1~mm change in water vapor results in a 0.01 ppm effect on our measurement signal. We can therefore safely ignore the impact of these lines.

\subsubsection{Scintillation}\label{sec:scint}
Turbulence in the upper atmosphere causes spatial intensity fluctuations due to changes in the index of refraction of warmer and cooler pockets of air that act to focus and defocus the wavefront. This causes speckling on the pupil image plane that imparts added random noise on the photometric measurements as these speckles change in intensity over time. This noise is called scintillation noise and for large aperture telescopes the magnitude of this noise term depends largely on the telescope aperture size, the exposure time of the observation, and the airmass of the measurement \citep{scint_osborn15}. 

A benefit of our oxyometer design is that we can simultaneously observe a star in both of our narrow bands. If our instrument produces exact copies of the telescope pupil for each of our two bands (for example, splitting the collimated telescope light with a beam splitter), then scintillation noise will be the same in each image as long as the effect does not evolve significantly over the 1~nm wavelength separation of our two bands. However, if our instrument operates by splitting the image of the star into two regions at the pupil level (this is done in \S \ref{sec:onsky} with custom-cut wedge prisms) then each of the two beams will experience different speckling and therefore we can expect scintillation noise to impact each band differently. For both cases we do not expect scintillation noise to be large, especially in the case of a large telescope, which will be needed to make an oxygen measurement. In the detection of potassium by OSIRIS, \cite{osiris} found that with their 10-m aperture the scintillation noise was negligibly affecting their measurements and their multi-band measurements were not performed simultaneously. Additionally, with wavefront correction an adaptive optics (AO) system can correct wavefront distortions in real time to reduce scintillation noise by an order of magnitude \citep{osborn_AO}. We anticipate that future large aperture telescopes will employ sophisticated AO systems that operate at the red optical wavelengths of the oxygen bandhead.

The dependence of scintillation noise on wavelength is a concern for small telescopes of around tens of centimeters in diameter and smaller because of the similar spatial scale of the speckles. We are only considering telescopes much larger, where the wavelength dependence is negligible. A second wavelength dependent effect occurs due to the delay of the scintillation noise due to chromatic dispersion of the atmosphere which worsens as observations move away from zenith \citep{scint_Dravins97,scint_osborn15}. Although a larger telescope aperture decreases the amplitude of the scintillation noise, this chromatic effect is independent of telescope size and increases with larger wavelength separations between the two photometric bands. Because the separation in our two bandpasses is $\sim$1~nm and at red optical wavelengths, we expect the delay in scintillation noise between our two bands to be negligible. Additionally, the $\sim{15}$-minute integration times the oxyometer would likely employ will be much longer than the millisecond timescales of any potential chromatic delay.

Although we do not expect wavelength dependent effects, in the case of splitting the pupil image into two separate beams, this effectively splits the aperture size in half by reducing the area over which the speckles are averaged. For long exposures the scintillation noise goes as $D^{-4/3}$, where D is the telescope diameter \citep{scint_osborn15}. For a 10-m diameter aperture this means the scintillation noise would go up by a factor of 2.5 if the pupil is split in half. For a 7th magnitude star, a 15-minute integration, and a 10-m telescope at a site like Mauna Kea, the scintillation noise as predicted by Young's approximation and adopting the pre-factor presented in Equation 7 of \cite{scint_osborn15} is about 20\% of the photon noise in one of our 0.3 nm width bands for a zenith angle of 30 degrees. This would rise to 60\% if we effectively split the telescope diameter by a factor of two and becomes a significant contribution to the overall noise budget. We considered how including scintillation noise in our simulations would impact the results shown in Figure \ref{fig:SNR} for the case where we must split the telescope pupil in half and found it increased the required transits required to reach 3$\sigma$ by around 10-15 for the GMT, TMT, and E-ELT and an increase of 30 transits for the smaller telescopes. When assuming a factor of 10 reduction in scintillation noise due to adaptive optics, the effects of scintillation only slightly reduce the transits required for the easiest targets. Without the ability of future AO systems to operate at a $\lambda$= 760nm with a $\sim$10\% noise reduction level, an oxyometer design that splits the incoming telescope beam and incurs scintillation noise will be much less feasible.

\subsubsection{Airglow}
A last concern for telluric atmospheric noise is airglow. Airglow is a term used by astronomers to describe any light emitted from the atmosphere. There are several mechanisms that can cause light emission in Earth's atmosphere, but the most common source is due to hydroxyl (OH) vibrational transitions \citep{OH}. Sky emission lines are known to vary in time and spatially across the sky and additionally brighten with airmass \citep{OHnightglowVariations,Gao_OHvariation}. To check for potential contamination due to overlapping sky emission lines, we check the catalogs of \cite{OHemiss_Hanuschik2003} and \cite{OHemiss_Osterbrock96} for sky emission in the regions we are interested in. These catalogs are derived empirically from ESO's UVES instrument on the 8.2 m UT2 of the VLT array and the HIRES instrument on the Keck 10 meter telescope, respectively. Most of the sky emission lines are due to OH, but there are also regions of O$_2$ emission and other common lines such as sodium and [O I]. From these catalogs we find that there are two regions free of sky emission around the oxygen 760 nm bandhead, which was what originally guided our placement of the two oxyometer bands. Using the \cite{OHemiss_Hanuschik2003} catalog, we further investigate the OH emission line at 758.6 nm. The FWHM of this line is 0.02 nm, which is narrow enough that overlap into the wings of our filters is not a concern. However, simultaneous monitoring of sky emission using moderate-resolution spectroscopy of sky regions around the target stars should help to strongly constrain the impact of variable OH emission on our measurements. 
 
\subsection{Stellar color variation}\label{sec:stelnoise}
Another challenge in transmission spectroscopy is the color variability of the host star across its disk. Variations in the color of a target star could obscure the signal entirely or could potentially be confused for a color change due to the exoplanet's atmosphere. Chromatic effects related to the star can be due to spots and faculae that are rotating in and out of view, limb darkening, or a wide range of sources of intrinsic stellar variability. If the transiting planet passes in front of a stellar surface feature of differing temperature, this will change the effective spectrum of the star and no longer match what was assumed for the out-of-transit baseline stellar spectrum. These color effects will propagate to what is derived for the exoplanet. This effect has been found to explain features in transmission spectra of the TRAPPIST-1 system observed by HST \citep{zhang_sunspots}. Though observations of the stellar light curve can attempt to confirm the presence of spots on the host star, it is difficult to predict the effect due to the many possible star spot and faculae configurations in an exoplanet's projected path across the star \citep{morris_trappist_spots}. Ignoring center to limb darkening and assuming a uniform stellar disk will also result in a systematic offset in the light curve as a function of wavelength \citep{howarth_center_to_limb,neilson_cent_to_limb}. We do not calculate the effect in our signal due to the wavelength variation of the center to limb darkening although how this effect manifests itself at high spectral resolutions should be quantified in future work. 

To understand the effect of stellar contamination due to stellar activity, we look to \cite{rackham_stellarvariability} who modeled the extent of the effect of a heterogeneous stellar photosphere on the exoplanet transit depth as a function of wavelength for M dwarfs considering realistic spot and faculae coverage. In this study the authors found that the true transit depth can differ by as much as 10\% from what is measured at the wavelength of the oxygen feature for early type M dwarfs with Solar-like spots. This is a large concern for estimating planetary sizes and deducing exoplanet spectra over a large wavelength range. To determine the effects of the contamination on our oxyometer wavelength bands, we utilize the results of \cite{rackham_stellarvariability} in their high resolution form and find that the narrow, dense stellar lines in M-dwarfs can be just as severe over our two oxyometer bands causing several parts per million effects in changing the transit depth signal. For M5V stars and larger, the variability in the ratio of our two bands due to the presence of giant spots is less than 5\% of the signal, while if we assume Solar-like spots the contamination effect is much more drastic and can double the expected atmospheric signal. Stellar activity is therefore certainly an important effect that must be considered in characterizing exoplanets and must be addressed carefully for each system. Further study of this important source of systematic uncertainty is required, especially beyond the assumption that spots and faculae can be modeled as cooler or hotter stars. In practice, time resolved transit photometry can help discern if a nonuniform stellar disk is affecting the measurements by looking for bumps in the light curve trough (see \citealt{morris_starspot,morris_trappist_spots,starspot_example}). Nonetheless, stellar contamination is an active area of investigation in the field of exoplanet atmospheric characterization and quantifying the extent of contamination will be important for validating future detections.

\subsection{Results}

\begin{table}
\centering
\begin{tabular}{lcc}
\hline
Telescope & Diameter & Effective Area \\ \hline \hline
Gran Telescopio Canarias (GTC) & 10 m & 74 m$^2$ \\ 
15 m & 15 m & 177 m$^2$ \\ 
Giant Magellan Telescope (GMT) & 25.5 m  & 368 m$^2$ \\ 
Thirty Meter Telescope (TMT) & 30 m & 655 m$^2$ \\ 
European Extremely Large Telescope (E-ELT) & 39 m & 978 m$^2$ \\ 
\end{tabular}
\caption{Telescope diameters and their effective areas for the observatories used in the calculations presented in Figure \ref{fig:SNR}.}
\label{tab:diameters}
\end{table}

\begin{figure*}
	\includegraphics[width=\linewidth]{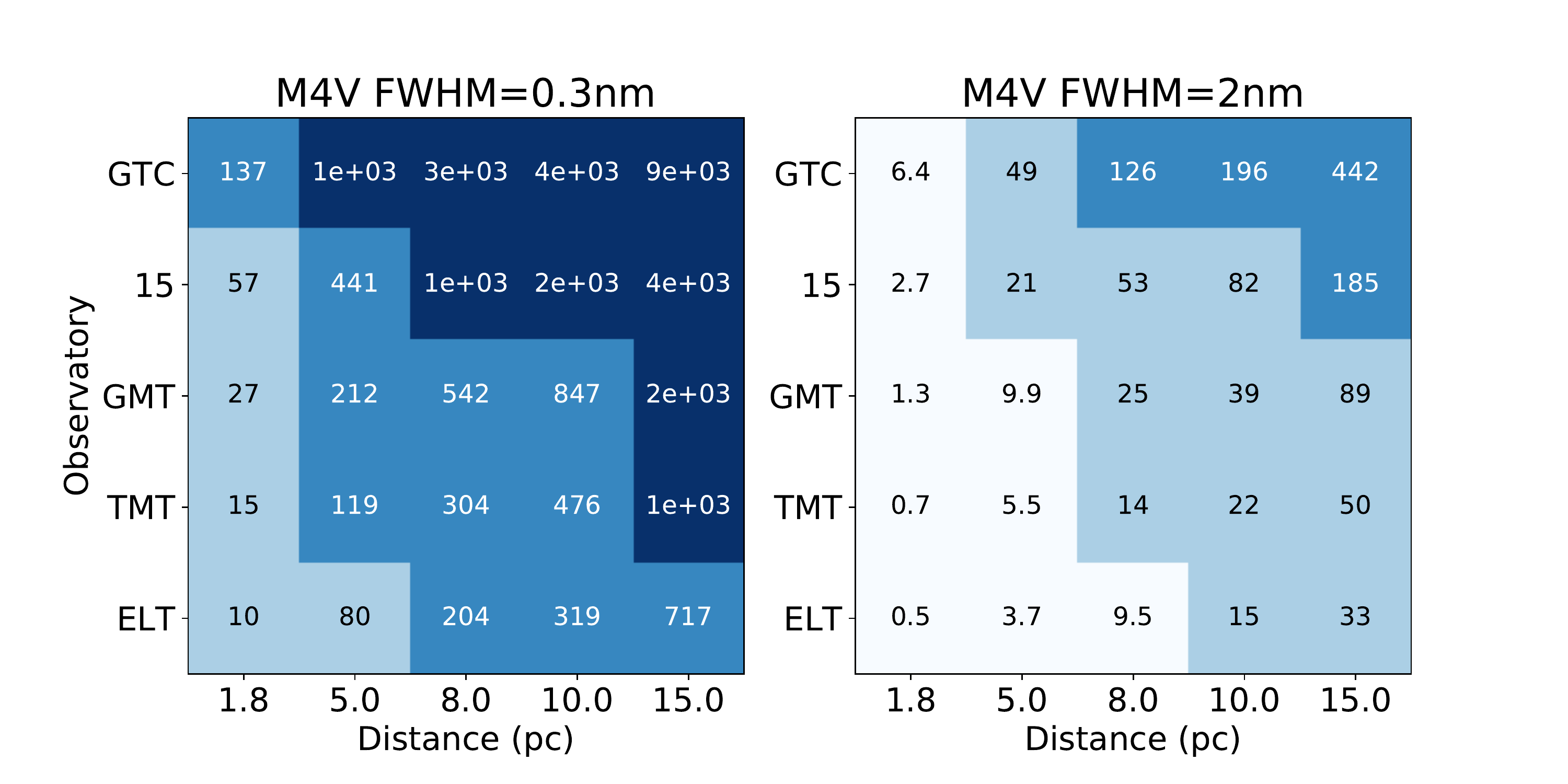}
    \caption{The number of transits required to reach a 3$\sigma$ detection of the oxygen 760 nm band for an Earth-twin orbiting an M4V star with $v_r$ of the M4V equal to -110 km/s, which is similar to Barnard's star. We present two cases: one for our proposed instrument with a filter bandpass of 0.3~nm FWHM (left) and the number of transits required for the case of a 2~nm FWHM filter overlapping the full oxygen 760 nm bandhead (right). We assume no 50\% beam split and a total instrument throughput $\eta$=50\%. Ground based ELTs will be able to target M4V stars and later if they are closer than 10 pc. The increased bandpass helps smaller telescopes reach fainter targets.}
    \label{fig:SNR}
\end{figure*}

\begin{figure*}
	\includegraphics[width=\linewidth]{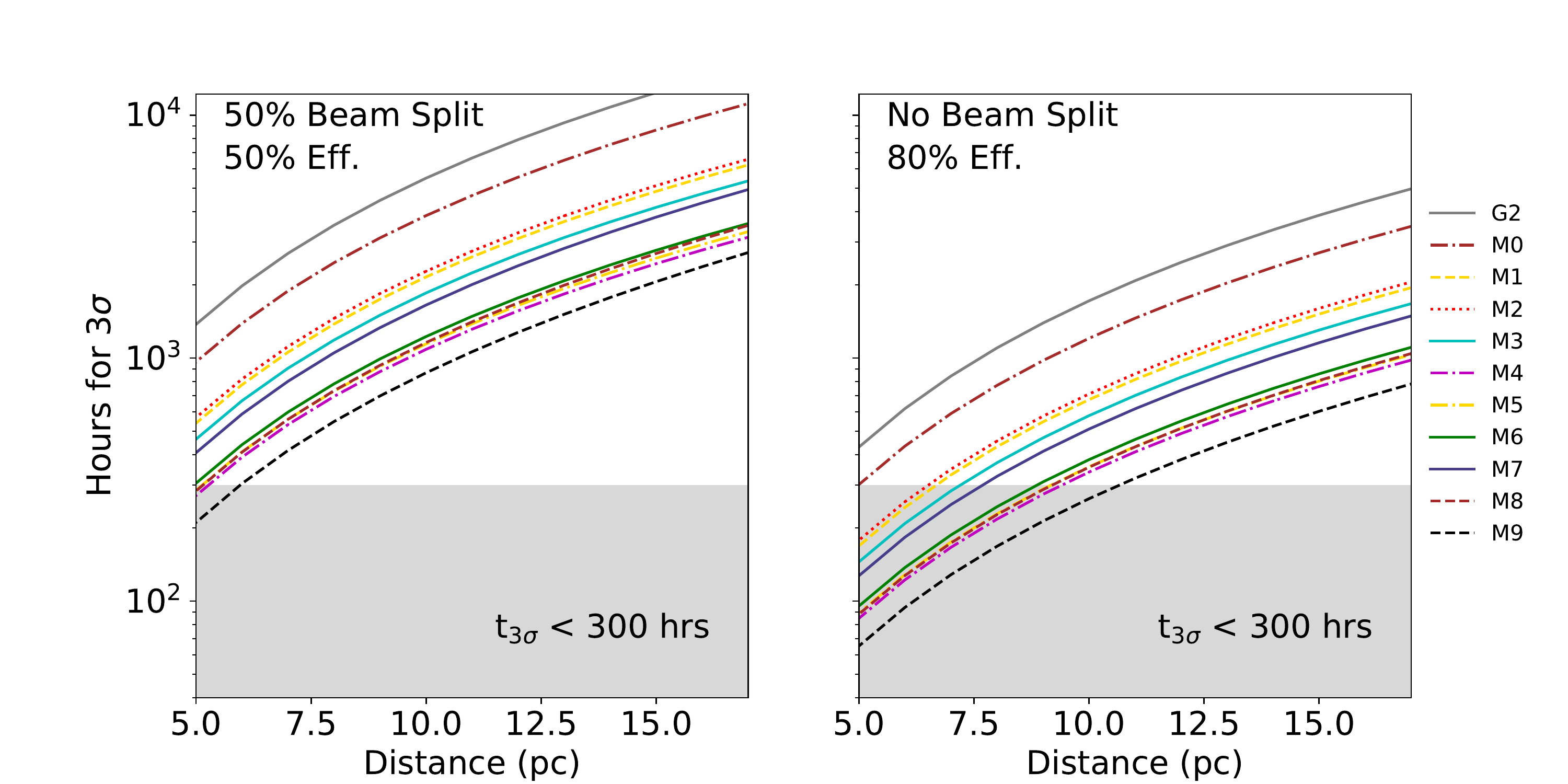}
    \caption{Observation time required to reach a 3$\sigma$ detection of oxygen for an Earth-twin around various host stars at different distances using an oxyometer on the E-ELT. We present two cases; on the left we show the case of an oxyometer with $\eta$=25\% which includes a 50\% beam split and additional 50\% CCD efficiency and telescope throughput and on the right we assume a more optimistic oxyometer that does not split the beam in half and has a total throughput of $\eta$=80\%. The easiest targets over all distances are M9V and M4V stars. For larger stars, the decrease in signal due to the increase in stellar radius in addition to the effects of refraction increase the time required to reach 3$\sigma$.}
    \label{fig:hours_snr3}
\end{figure*}

In Figure \ref{fig:SNR} on the left we show the number of transits required for an oxygen detection in an Earth-like planet orbiting an M4V star using a ground-based oxyometer assuming a range of stellar distances and effective aperture areas corresponding to the following telescopes: the GTC, GMT, TMT, E-ELT and a hypothetical 15 m aperture telescope. We list the values for effective diameter and collecting area we use for the calculations in Table \ref{tab:diameters}. The SNR was calculated assuming an M4 dwarf with a radial velocity of -110~km/s, similar to Barnard's star which is 1.8 pc away. For this star, the transit duration for an exoplanet in the habitable zone is 2.06 hours so each transit would require 2.06 hours of observing the system in transit and 2 hours of observing the system out of transit. On the right panel in Figure \ref{fig:SNR}, we perform the same calculation but assume a larger filter width. As was discussed in the previous section, the maximum oxygen signal for this system remains approximately the same even if the filter overlaps the entire oxygen bandhead. For ground based calculations with an oxyometer where the FWHM of the filter is 0.3~nm, the E-ELT at an effective diameter of 35.3 m should be capable of making a detections for select M4V stars at a distance closer than 8 pc. The slightly smaller TMT with a 28.9 m effective diameter may be able to characterize similar systems out to 5 pc, and for the case of an Earth-like exoplanet orbiting Barnard's star, which is 1.8 pc away from Earth, a ground-based detection with a telescope as small as 15~m in diameter is practical. In the case of a wider bandpass, telescopes as small as 10 m would be able to characterize systems closer than 5 pc. This means that a future space telescope like LUVOIR, a current Decadal survey mission concept, that is equipped with an instrument capable of at least R$\sim$400 spectrophotometric measurements operating near the photon limit would be able to detect oxygen in an Earth-like planet around late type M dwarfs (assuming a $\sim$10 m aperture). For low resolution atmospheric oxygen studies of upcoming TESS targets (with no restriction on the system's radial velocity) a space-based telescope like LUVOIR with a science driven instrument design will be required. Otherwise, only the most accessible systems that are nearby with sufficiently high radial velocities to avoid telluric overlap will be feasible targets by ground-based ELTs paired with an oxyometer-like instrument.

In Figure \ref{fig:hours_snr3} we present our results for a range of stellar types and plot the observation hours required to reach a SNR of three as a function of stellar distance. The results of a lower throughput instrument that operates by performing a 50\% beam split are shown on the left, whereas a more optimistic, but still realistic, instrument is shown on the right for no beam split and a total instrument throughput of 80\%. Here, we assume the 0.3 nm width bandpass and take the maximum signal corresponding to when the host star is moving at least 175 km/s towards Earth. In both cases the targets requiring the fewest observing hours are M9 stars followed by M4, M8, M5, and M6 stars in order beginning with the least observing time necessary for a 3$\sigma$ detection. For the optimistic oxyometer design with an overall higher throughput, these late type M dwarfs at a distance of 10 pc still require from 265 to about 500 total observing hours, while at a distance of 5 pc the observing time ranges from 64 to 96 hours. For the case of an oxyometer with 50\% beam split and 50\% efficiency, the easiest target at 5 pc, an M9 dwarf, requires 210 hours. This demonstrates the benefit of building a high throughput instrument with optimized mirror coatings and highly efficient detectors. 

Previous work has investigated the observing time required for proposed instruments on the GMT, TMT, and E-ELT to observe the 760 nm oxygen signal, but utilizing a different method for detection. In \cite{rodler14} and \cite{snellen_o2} they investigate the capabilities of high resolution (R\textgreater100,000) spectroscopy to resolve the exoplanetary oxygen lines to perform a cross correlation with a template oxygen spectrum in order to search for the signal that will be Doppler shifted according to Earth's barycentric velocity. \cite{rodler14} conclude that with this method on the E-ELT with a UVES-like instrument design, M dwarfs of types later than M4 will be the easiest targets with observing times to reach a 3$\sigma$ detection ranging from 29 to 43 hours for late-type M dwarfs 5 pc away. The high resolution cross correlation (HRCC) technique utilizes the many oxygen lines at changing planetary orbital velocities to detect and validate a signal much below the photon noise of the data making it a very powerful method. Despite this, the observing time required for obtaining a lower resolution transit spectrum detection is on a similar scale. Since the HRCC measurement is sensitive to the atmospheric spectrum's line shapes, which are generated in the upper atmosphere, while lower resolution transit spectrophotometry is more sensitive to the baseline change in the planet radius in and out of transit, which is dominated by absorption in the lower atmosphere, combining measurements with these two methods would further constrain atmospheric models \citep{HRCCandLDS}.

\section{Sample Size of Viable Targets} \label{sec:targetsize}
In the case of ground-based observing with the oxyometer, it is necessary to ask how many viable targets we can expect provided the limitation that our target stars must have a sufficient negative radial velocity, $v_r$, to shift an exoplanet signal from the telluric oxygen lines. Additionally, since the targets amenable to characterization must be within 10 pc of Earth, this further reduces the population. The FWHM of the oxyometer's `on' wavelength bin of 0.3 nm set the minimum $v_r$ of the target stars that are observable by our oxyometer design. Ultimately, we want to maximize the oxygen absorption signal overlapping the filter bandpass, while also completely avoiding telluric oxygen absorption lines. Our filter must then be at least half a width (0.15 nm) away from the telluric oxygen bandhead. Stars with small negative radial velocities will result in lower overlap between the filter and the oxygen signal from the exoplanet atmosphere and therefore will be more difficult to detect. Figure \ref{fig:f_v_vel} shows how the signal, $\mathcal{S}$, varies as a function of the radial velocity of systems with M4V and M6V host stars. Which stars are observable depends on the final instrument sensitivity, type and distance of the star, as was shown in Figure \ref{fig:hours_snr3}. We decide the cut-off velocity for systems that are potential candidates for being observed with the oxyometer design proposed in this paper to be -75~km/s, where the signal drops to around 2 ppm for an M4V host star. This limit will change for the specific planet and host star properties; however, it provides a limit for which we can estimate the sample size of potentially accessible targets for ground-based characterization with our oxyometer.

To estimate the fraction of M dwarfs that we can expect to have $v_r$\textless-75 km/s, we use the fifth data release of the RAVE survey. The RAVE survey focused on bright stars in the southern hemisphere. Since stars in RAVE DR5 are randomly selected for observation if they pass their photometric cut, the sample does not contain kinematic biases \citep{rave5}. We do observe that the fraction of stars with radial velocities faster than -75~km/s increases with lower mass stars. Therefore, to represent the population most appropriate for the stars we are interested in, we take a subsample of stars from the RAVE catalog that have similar photometric properties to the TESS M dwarfs \citep{TESSoccurence}. To create a subsample of stars that are similar to the temperatures and distances of TESS M dwarfs, while still having a statistically useful sample, we consider distances less than 200~pc and temperatures less than 4000~K. This reduces the $\sim$458,000 unique stars from the full DR5 RAVE catalog to about 2500 stars. Of these 2500 stars, 5\% have $v_r$ between -600 and -75~km/s. 

From the results of \S \ref{sec:noise} we found that the optimal targets have stellar types later than M4V with distances closer than 10 pc. A query of the all-sky bright M dwarf catalog presented in \cite{BrightMDwarfCatalog} shows 114 stars for M stars later than M4V and 29 for types later than M6V, if restricted to distances less than 10 pc. The authors estimate their catalog is complete to 75\% for J \textless 10. 
Applying the expected fraction of high radial velocity stars to the M dwarf all-sky counts, we estimate seven or eight M4V or cooler stars will be near enough and have high enough radial velocities for characterization in the case that they have transiting Earth-like companions. 

Of these stars, we must now consider the probability that they will have a transiting planet. We therefore apply the transit probability, $P_{\mathrm{tr}}$ = $R_s/a$, where $a$ is the orbital distance. For an M4 star with a planet in the habitable zone at 0.05 AU, $P_\mathrm{tr}$ = 0.02 leaving the number of stars accessible to characterization at around 0.1. This is a lower limit since the probability for transit would be slightly higher if considering later type stars. While the odds for a star in this sample hosting a planet make the likelihood of finding an ideal target marginal, the significant possibility of finding even just one exoplanet ideal for oxygen characterization is exciting. Additionally, Barnard's star at 1.8 pc away is a type M4V moving towards Earth at 110 km/s \citep{barnardstar}, making it an ideal host star for ground-based characterization with our oxyometer if it had a transiting companion. 

The small number of expected targets due to avoiding telluric oxygen motivates ways to increase the sample size by either constraining telluric variability such that a bandpass overlapping Earth's absorption bands is feasible or by installing an oxyometer-like instrument on a large (\textgreater 10m) space based telescope.

\section{Building an Oxyometer}\label{sec:lab}
To demonstrate an instrument that achieves the same wavelength resolution and band spacing of the proposed oxyometer, we first set up our instrument design on an optical bench and perform various photometric tests. We use an ultra narrowband interference filter centered at 607.3 nm with a FWHM of 0.33 nm purchased from Alluxa. Using this setup we show that this instrument can operate near the photon noise limit and demonstrate a detection of a simulated 50 ppm signal. 

\begin{figure*}
    \centering
    \includegraphics[width=\linewidth]{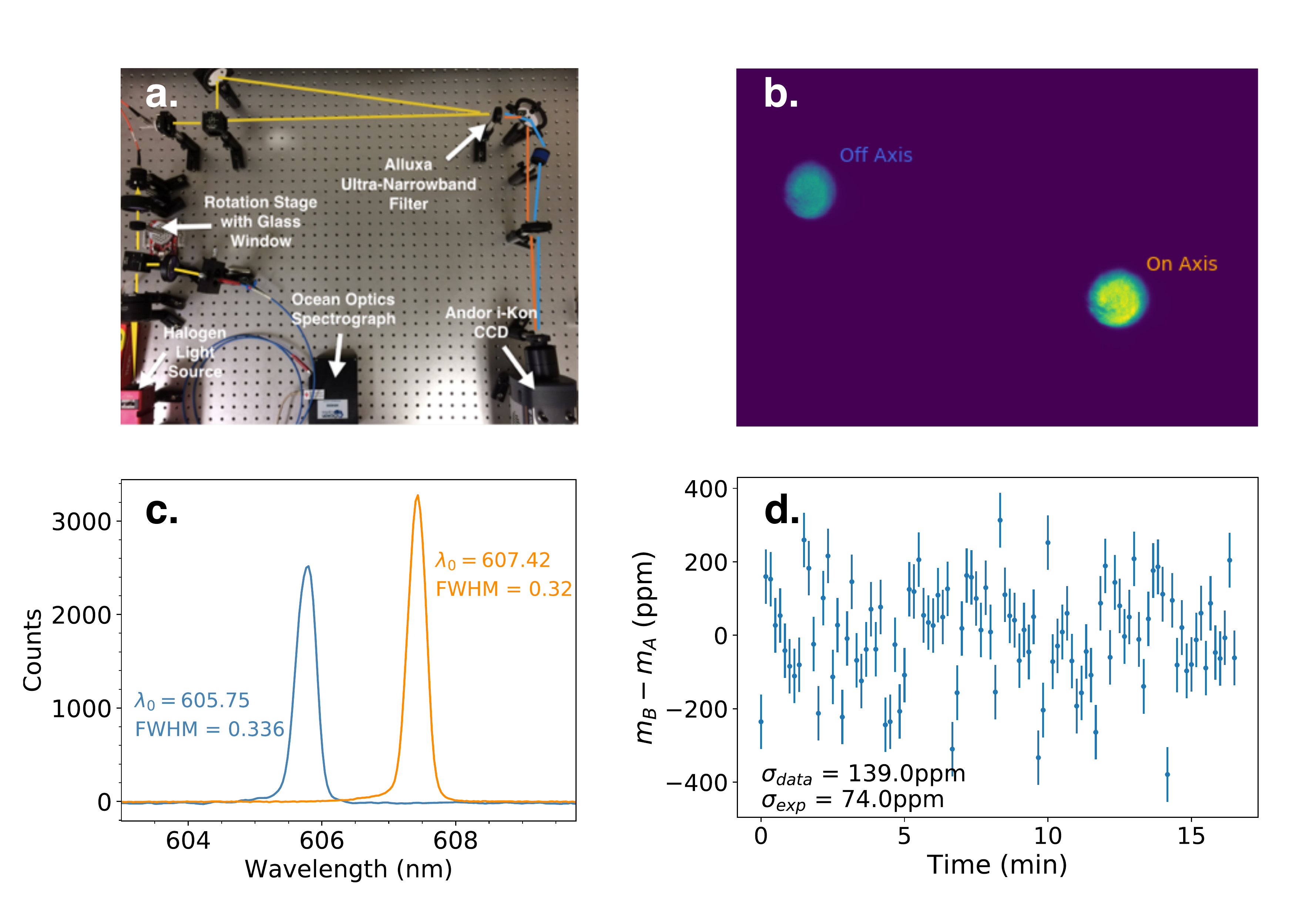}
    \caption{Images of the first oxyometer setup and data. (a) The optical setup of the instrument. (b) Example image of the two beams. (c) Spectra of the on-axis and off-axis output beams with central wavelengths and FWHM values labeled in units of nanometers. (d) Time series of the data taken with the Andor IKON. }
    \label{fig:lab_setup}
\end{figure*}

\subsection{Optical Layout}
We show the optical table layout for the first oxyometer prototype in panel (a) of Figure \ref{fig:lab_setup}. For a light source we use a halogen lamp\footnote{Originally we were using a supercontinuum laser as the light source for our tests, but we found that high frequency intensity variations dominated the measurement.}. The collimated light first passes through a 92/8 beam splitter. The 8\% side passes to an Ocean Optics USB4000\footnote{https://oceanoptics.com/} spectrograph to independently monitor the lamp's spectrum to track temperature variations that would affect the color of our source beam. The remaining 92\% of the light continues through a 1 mm thick uncoated glass window made of N-BK7 material mounted on a rotation stage that allows the glass plate to be tilted at small angles (down to $\pm$0.0165 degrees) relative to the collimated beam. After passing the glass window, the beam is focused onto an optical fiber cable whose other end is coupled to a collimator shown in the top left corner in Figure \ref{fig:lab_setup}a. The collimated light passes through a 50/50 beam splitter with one beam passing directly through the Alluxa filter while the other beam is reflected off a mirror and directed such that it passes through the Alluxa filter at 8 degrees from normal incidence. This difference in incidence angle between the on-axis and off-axis beams causes them to experience effective passbands that differ in wavelength. The off-axis beam passes through a wedge prism to reduce its angle with respect to the on-axis beam from 8 degrees to 2 degrees before they are focused. A partially reduced angle of the off-axis beam is necessary so that both beams can fit onto one CCD chip while still focusing at different physical positions such that the beams do not overlap one another. At the focal plane we include a 40 nm wide bandpass filter that helps to reduce contamination from background light. An example image of what the two beams look like when imaged with the described setup is shown in Figure \ref{fig:lab_setup}b. Additionally, in Figure \ref{fig:lab_setup}c we plot the observed spectrum of the two beams taken with the Ocean Optics spectrograph.

The first segment of the optical setup, which includes the 1 mm thick glass window, serves to induce a color-dependent absorption signal on the input beam when the window is tilted slightly. This glass window was selected based on the shallow wavelength dependence of the transmissivity of the glass. We observe that every 1 degree tilt of the window causes a 650 $\mu$mag difference in the magnitudes of the on-axis and off-axis beams. This glass window is therefore a precise way to impart a repeatable color difference on the output signal with a known and adjustable magnitude in order to simulate chromatic transient events.  

It is important to note that, due to cost, we are not using a custom-made filter centered at the wavelengths assumed for the calculations made in the previous sections. Instead we purchased a previously manufactured Alluxa filter centered at 607.3 nm with a FWHM of 0.33 nm. We chose this filter because of its narrow bandpass that is similar to the 0.3 nm FWHM we required for our theoretical oxyometer instrument, and additionally because its central wavelength avoids telluric lines and airglow emission, making it possible to use for on-sky tests that match the conditions of the assumed oxygen-centered filter used in the previous sections' calculations. 

\subsection{Results of Photometric Tests}
Measurements were taken with the bench-top prototype with the goal of testing the photometric performance. Using a halogen lamp and an Andor IKon-M CCD camera\footnote{We used was the Andor IKon-M 934 deep depletion CCD operating at -60 K and a gain of 5.2 electrons/ADU with a grid of 1024x1024 13$\mu$m square pixels.}, we imaged the two beams for an integration time of 4~s at a cadence of one image every 10~s for a total duration of 15 min. An average of dark frames of the same exposure time were subtracted off each image and aperture photometry was performed on each beam with sky-subtraction methods also implemented to remove any remaining background flux due to scattered light in the system. This was important because we defocused the beams to spread the light around a larger number of pixels to average over flat-field effects. Our photometric aperture was therefore quite large ($\sim$25,000 pixels) and so even a small per-pixel background level could translate to a significant total background flux, which was mostly due to scattered light from the halogen lamp and therefore would not be present in the dark frame. We compute the flux ratio of each beam, $F_b/F_a$, where $F$ is the flux measured in each band in units of photons and we follow the convention of $a$ representing the on-axis band and $b$ representing the off-axis band. We plot the magnitude difference, $m_b - m_a = -2.5\log{F_b/F_a}$. In this case the on-axis band is brighter, so $m_b - m_a$ is positive.

To calculate the expected noise of our measurements, we use Equation \ref{eq:CCDeq} to calculate the photometric error in the flux of each beam after converting to photons from electrons using the measured gain of the camera, which is 0.37 electrons per ADU. We similarly propagate the noise in each beam to the flux ratio as was done in Equation \ref{eq:s_err_in} and further propagate this to error in the the magnitude difference, $\sigma_\textrm{m}$:

\begin{equation}
\sigma_\textrm{m} = \frac{2.5}{\ln{10}}\frac{F_\mathrm{b}}{F_\mathrm{a}} \sqrt{ \frac{\sigma_a^2}{F_{\mathrm{a}}^2} +  \frac{\sigma_b^2}{F_{\mathrm{b}}^2}} \:\:.
\label{eq:sig_rat}
\end{equation}

\noindent We show the results of the data collected with the IKon camera in panel (d) in the bottom right of Figure \ref{fig:lab_setup}, which shows the expected error bars calculated using Equation \ref{eq:sig_rat}. The standard deviation of the data shows a scatter of 139.0 ppm while the error bars show the predicted noise values of 74 ppm. During the data acquisition the halogen lamp can change temperature and heat up as it is left on for longer. This imparts a slowly varying trend to the time series of the ratio of the two beams that was confirmed independently by taking simultaneous spectra of the lamp. The segment shown here is after a smooth third order spline was removed form the flux ratio data based on a fit to the temporal variation of the lamp spectra. There appears to be some correlated noise in the data stream. It is not clear why the observed noise levels are about double what is expected, though residual flat-field errors coupled to subtle shifts in the spot positions on the CCD are a potential source of concern. 

\subsection{Detecting a Simulated Transit}

Using the same setup presented in the previous section, we wished to demonstrate the ability to detect a 50 ppm level color change in the light source. To create a change in flux between our two bands, we use the glass window mounted on a rotation stage (labeled in panel a of Figure \ref{fig:lab_setup}) that the halogen light passes through before it is coupled to the beam splitter and passes through the Alluxa filter. By tilting the glass window we can impose a chromatically dependent attenuation on the throughput beam. We determine the tilt angle corresponding to a 50 ppm drop in the flux ratio of our two bands by tilting the glass window in 0.2 degree steps. We find that the color change per degree was nearly linear and equal to 650 ppm per degree tilt. Therefore, to create a faux-transit 50 ppm signal, a 0.08 degree tilt to the glass window is required. The positioning error and repeatability of the ELL8 model rotation stage from Thorlabs\footnote{www.thorlabs.com} that the glass window is mounted on are 0.0165 degrees and 0.025 degrees, respectively. While these errors are somewhat large with respect to our target rotation, we can expect that the errors will average out over many iterations of cycling between the angled and not angled positions such that the resulting signal will be near 50 ppm. 

\begin{figure*}
\centering
\begin{minipage}{.48\textwidth}
  \centering
  \includegraphics[width=0.99\linewidth]{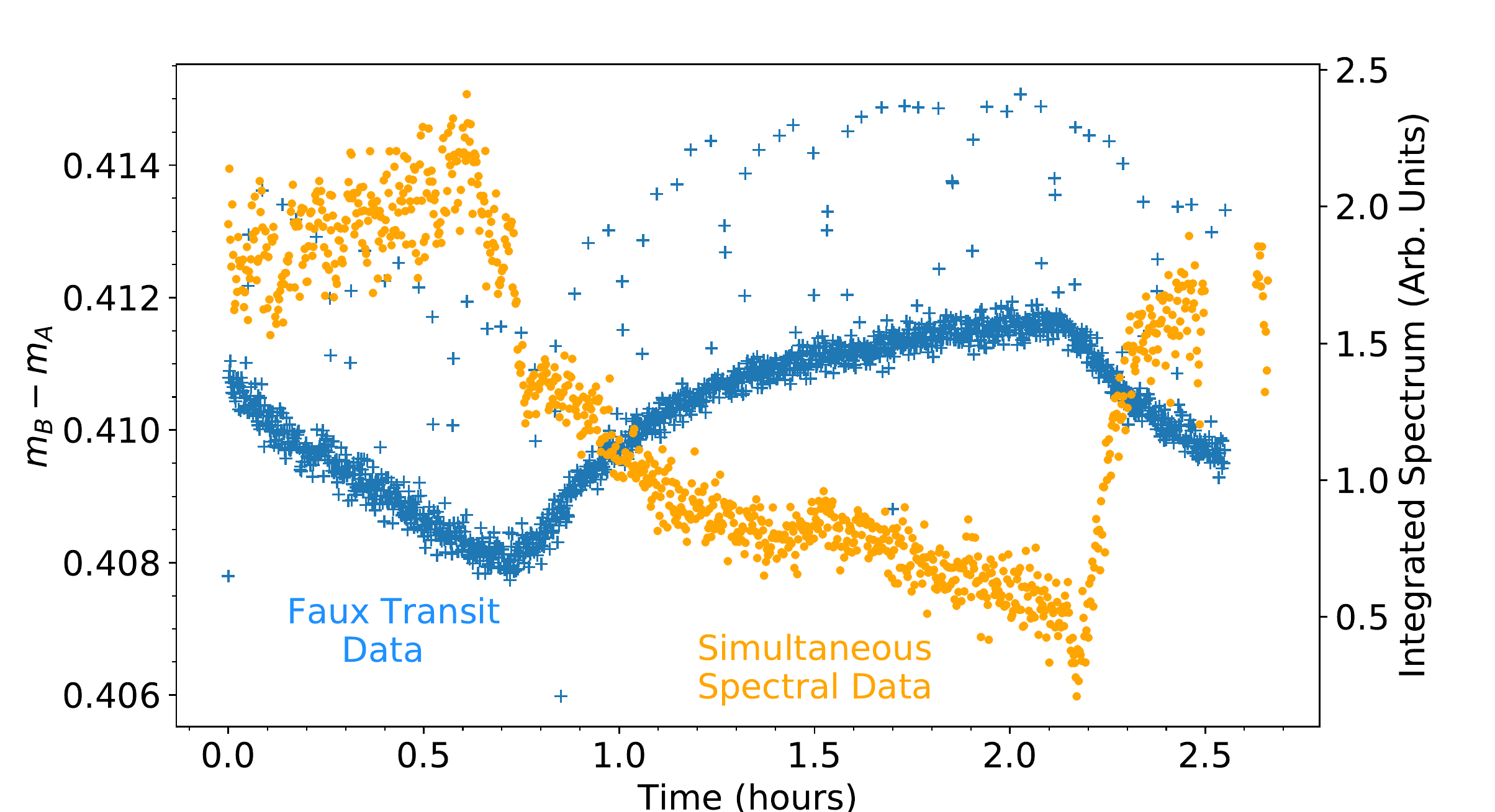}
  \caption{The flux ratio data stream for the simulated transit detection in blue and the simultaneous spectral data in orange. The spectra were integrated over the wavelength range from 604 nm to 615 nm and show an inverse correlation to the flux ratio from our two photometric bands at 606.5 nm and 607.3 nm. The blue points periodically deviating from the flux are where we tilted the glass window by 1 degree to mark the change in position between the no tilt state and the 0.08 degree tilt state. }
  \label{fig:spec_flux}
\end{minipage}
\hfill
\begin{minipage}{.48\textwidth}
  \centering
  \includegraphics[width=0.99\linewidth]{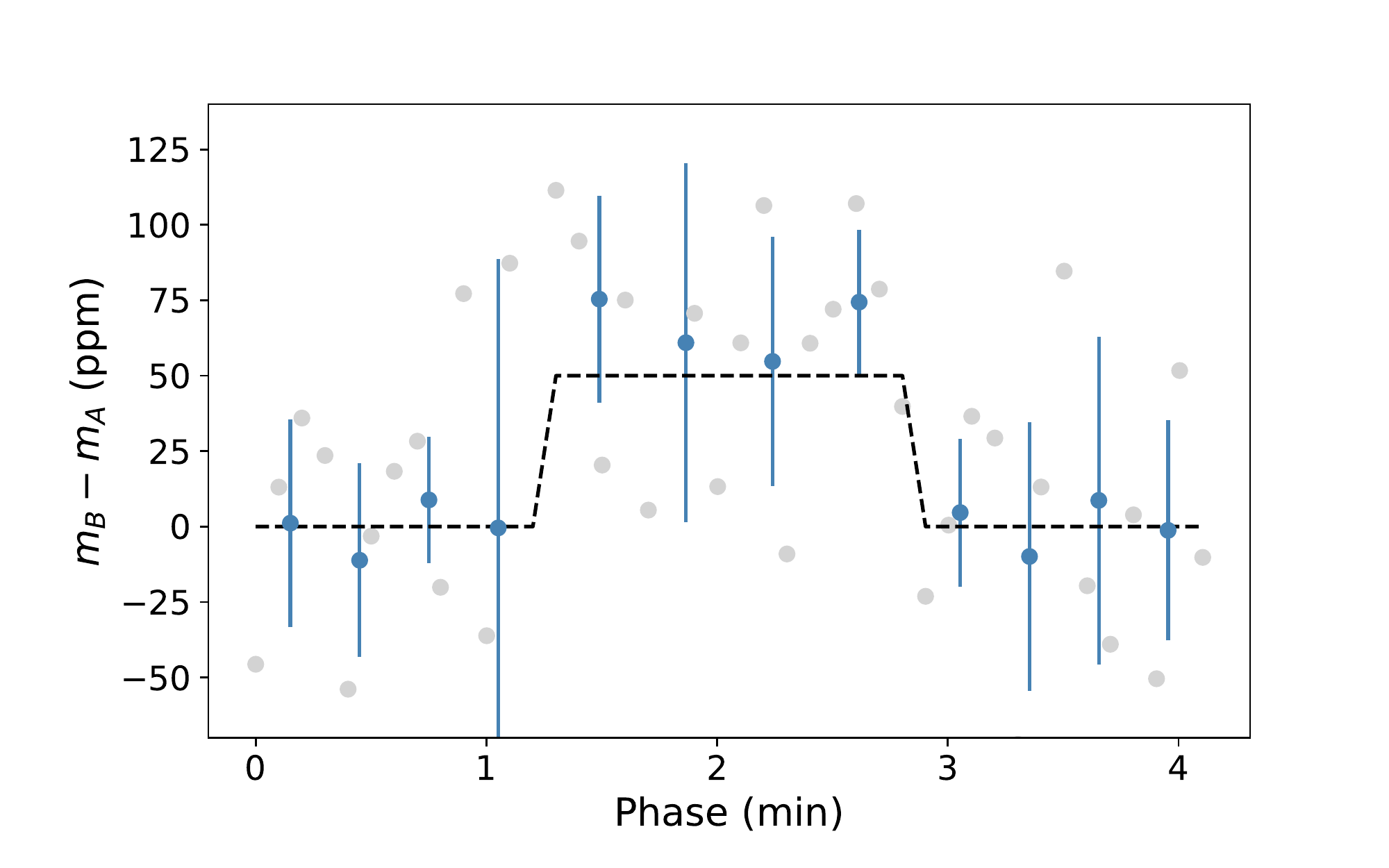}
  \caption{Simulated transit taken with the Andor iKon CCD by tilting a glass window 0.08 degrees to produce a 50 ppm color change between the two bands. The blue data points show the data binned in time and averaged for 12 `transits'. The gray points are the same data before binning, but after averaging the transit and the dashed line shows the expected signal.}
  \label{fig:transit}
  \end{minipage}
\end{figure*}

For the data collection, we continuously image the beams in three separate 51 min sequences. In that time we cycle between tilting the glass window 0.08 degrees and returning the window to the 0.0 degree position\footnote{Other attempts to create a small color change in our light source included sliding a glass window in and out of the path of the light source. The observed effect was several times larger than expected provided the transmission spectrum of the material, which suggested a 0.02\% difference in transmission between our two bands, possibly due to reflections off of the glass surface.}. Between each transition we also move the filter by a larger amount (1 degree) in order to mark in the data stream the time when a transition occurred. The halogen lamp also was changing in flux during these data runs, causing a varying continuum and also possibly a variation in the black-body spectrum of the lamp. In order to sufficiently sample this trend, we chose the cycle times to be 3 minutes of 0.1~s exposures every 5~s when the glass window is not tilted and 2 minutes of 0.1~s exposures when the glass window is tilted by 0.08 degrees. 

The final data stream of the flux ratio of the two bands was extracted in the same way as was described in the previous section. Similar to the previously described photometric run, variations occurring on 30 minute timescales with occasional dramatic dips were observed in the photometry. We analyzed the simultaneously recorded spectra by first subtracting a master dark frame from each spectrum and then integrating each spectrum between the wavelengths of 604 nm and 615 nm. This integrated flux serves as a measure of the halogen lamp's output. The time sequence for this integrated flux value of each spectrum is plotted in Figure \ref{fig:spec_flux} in which we also plot the time series of the flux ratio of our two photometric bands. It can be seen that as the output from the halogen lamp increases, the flux ratio decreases (becomes bluer). This is consistent with what we expect from blackbody emission: if the halogen lamp increases in temperature then the total integrated flux increases and the lamp becomes bluer, which results in a higher flux ratio, $F_a/F_b$ ($F_a$ is redder than $F_b$). This analysis of the simultaneously recorded spectra confirms that the changing temperature of the halogen lamp is causing the smooth variations in the flux ratio.

Although our simultaneous spectral information confirmed the source of the systematic variation in the flux ratio, the spectral data were too noisy to be able to detrend the flux ratio variations directly. We therefore adopt a different method to flatten the data. To do this, we fit a line to two consecutive `out-of-transit' segments of data. Then we subtract this line from the out-of-transit and in-transit data, taking care to only include half of the out-of-transit data on each side to avoid including some portions of the data twice. A few segments of data were excluded because there was substantial short-timescale structure in the halogen lamp output that could not be sufficiently corrected with a linear fit to the continuum. A final total of 12 simulated transits were averaged together. We plot these data in gray in Figure \ref{fig:transit} and overplot the same data after it has been binned. The expected 50 ppm signal is also plotted as a dashed black line. Despite the challenges related to the halogen lamp changing temperature, the signal is easily detected in our data set. An analysis of the variances of the data rejects the null hypothesis that the means are the same between the in and out of transit regions to a significance of p=1.2e-5. The result is consistent if we instead treat the `in transit' data points as the continuum and flatten the full data stream by using a linear fit to these data instead. In that case we measure the signal to be -46~ppm as opposed to 66~ppm in the reverse case. With this test we have demonstrated the ability of our oxyometer concept to detect small, chromatic transits that are similar in amplitude to a true transit.

\section{Compact Oxyometer for On-Sky Measurements}\label{sec:onsky}
For our second oxyometer prototype, we wished to devise a design that could be used on sky and that avoids the need for an optical fiber cable to couple the light from the telescope to our instrument. This is motivated by the desire to avoid modal noise that can occur in fibers, which can be highly chromatic \citep{fiber_redding}. The ideal instrument would be compact and could attach directly to the back of a telescope. We developed an instrument design that achieves this by incorporating an off-the-shelf lenses and custom-designed wedge prisms. Here, we discuss the optical design of the on-sky oxyometer and present the results of on-sky photometric tests.

\begin{figure*}
	\includegraphics[width=\linewidth]{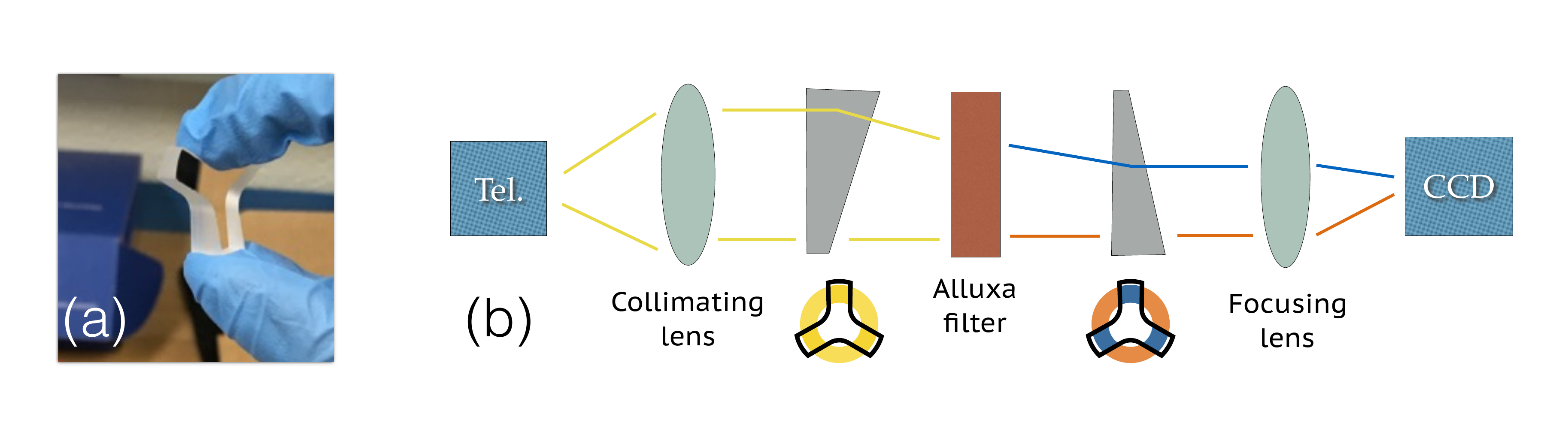}
    \caption{(a) One of a pair of custom made wedge prism optics. (b) An edge-on diagram depicting the setup of the wedge prism instrument. Light from the telescope is collimated, passes through the first wedge prism then the Alluxa filter. Photons that were angled by the first wedge prism will pass through the second wedge prism that serves to partially correct this angle, leaving a slight difference in the angle of incidence between the on and off-axis beams such that they focus on different regions of the CCD.}
    \label{fig:wedge}
\end{figure*}

\begin{figure*}
\centering
\begin{minipage}{.48\textwidth}
  \centering
  \includegraphics[width=0.99\linewidth]{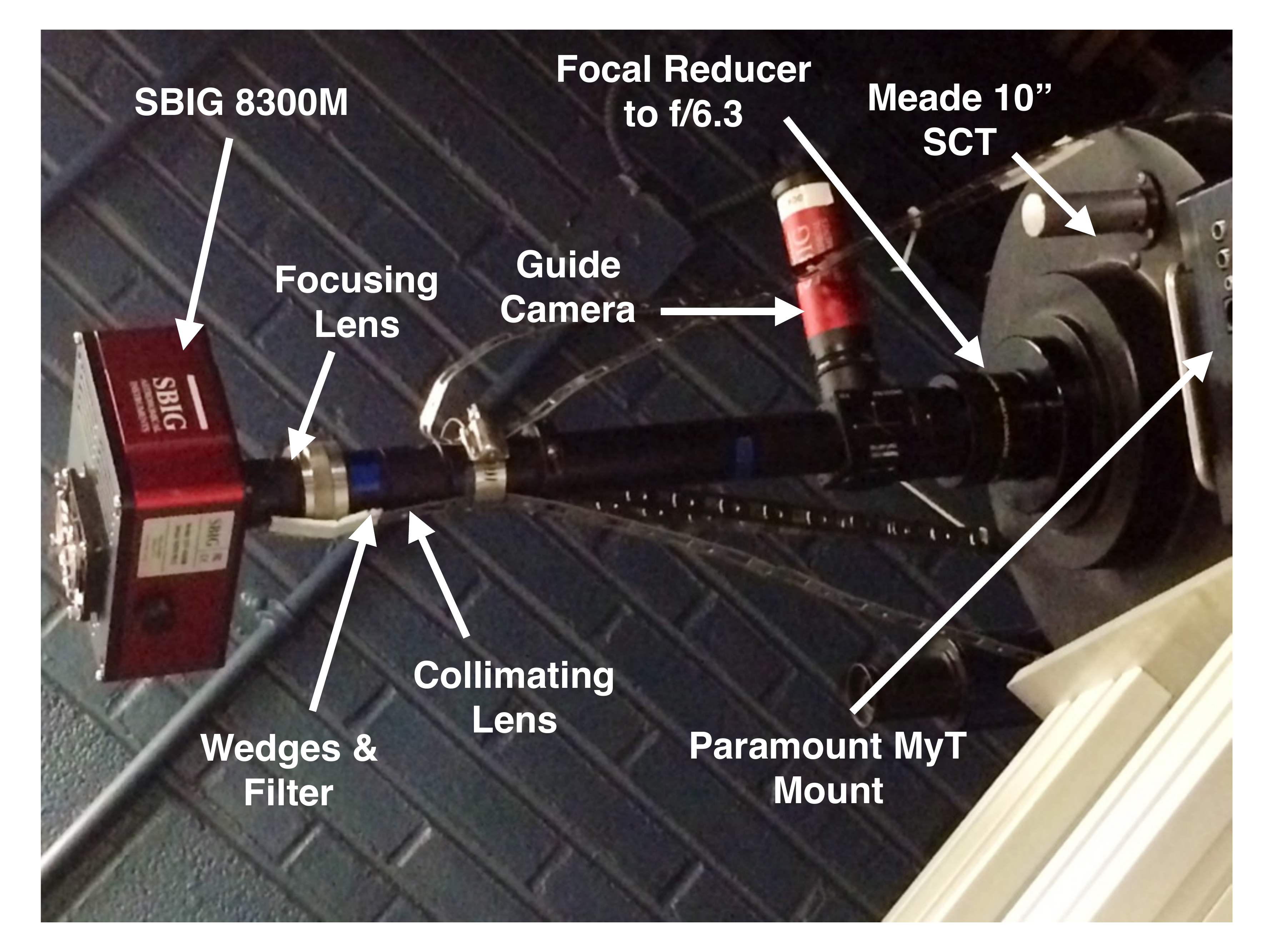}
    \caption{The wedge prism oxyometer setup attached to a 10" telescope at the University of Pennsylvania in Philadelphia. The telescope has a focal ratio of f/10 that we reduce to f/6.3 with a focal reducer. A beam splitter sends light to an ST-i for guiding before passing through the lenses, wedge prisms, and narrowband filter. On-sky tests were made using this setup. See text for more details.}
    \label{fig:onsky}
\end{minipage}
\hfill
\begin{minipage}{.48\textwidth}
  \centering
  \includegraphics[width=0.9\linewidth]{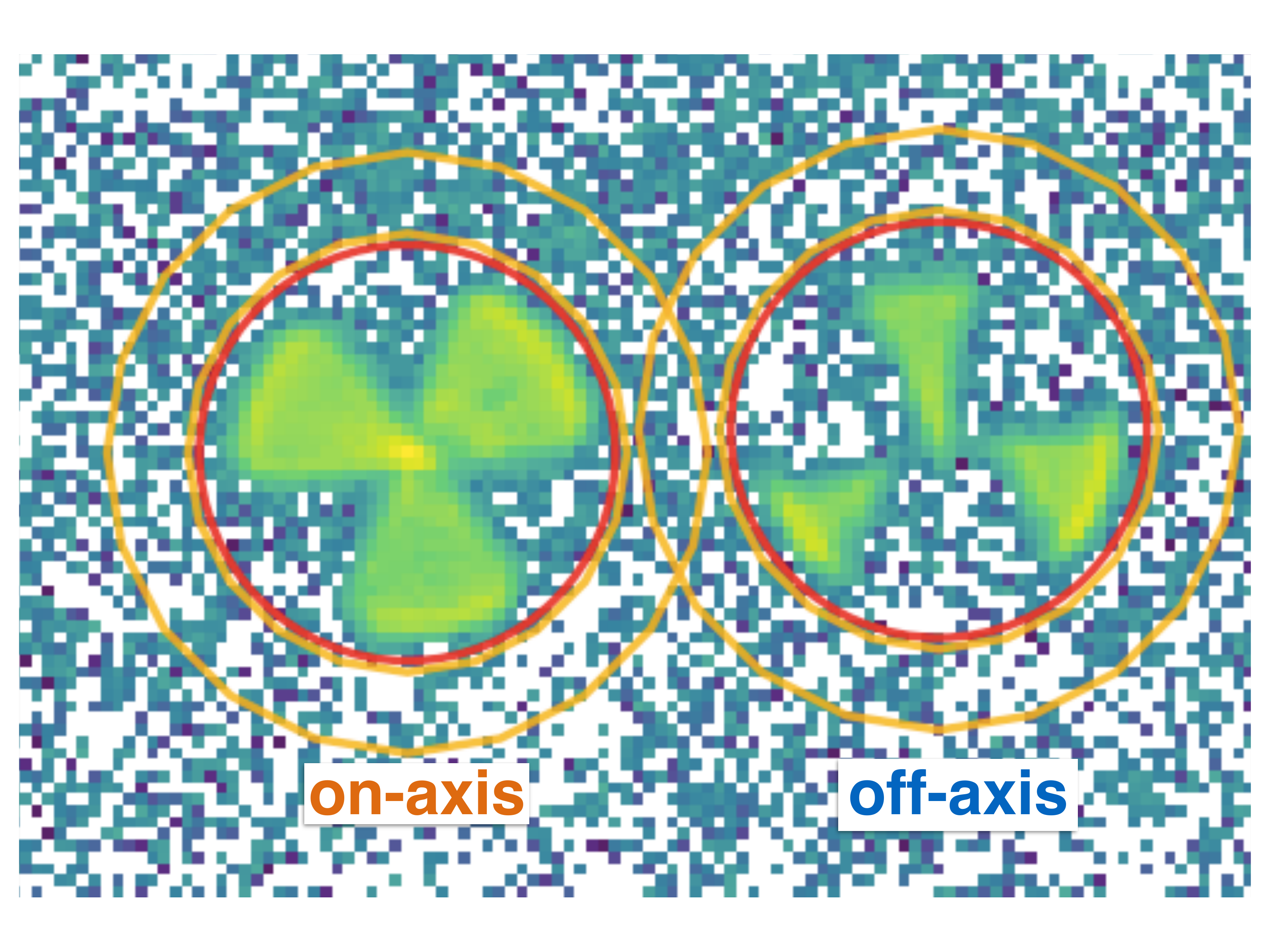}
    \caption{An example image of the beams taken with the SBIG 8300M. The beam on the left is the on-axis beam that passes through the empty space of the wedge prism, while the right-most spot is the off-axis beam that passes through the wedge prism and therefore experiences a bluer effected bandpass. }
    \label{fig:onskyspot}
  \end{minipage}
\end{figure*}

\subsection{Optical Design}
The compact, on-sky version of our oxyometer uses the same 607.3~nm Alluxa filter and additionally relies on wedge prisms that replace the mirror and beam splitter combination in the previous setup. These prisms serve to send half of our beam through the narrowband interference filter at an angle and the other half at normal incidence. These wedge prisms were custom-ordered to be cut into a triaxially symmetric `Y' shape. Figure \ref{fig:wedge} shows an image of one wedge prism in panel (a). The wedge prism is 1" in diameter and deviates incoming light at an angle of 6 degrees, slightly less than our previous design and therefore the resulting two effective bandpasses are slightly closer in central wavelength, but still minimally overlap. A cartoon drawing of the face of the prism post-cutting is shown in panel (b) in Figure \ref{fig:wedge} with the blue and orange shaded doughnut representing the resulting different effective wavelengths of an incident pupil image of a telescope with a central obstruction due to a secondary mirror. Panel (b) also shows how one of the wedges serves to split a 1" incoming collimated light source into two beams, one of which is angled by 6 degrees, while a second wedge serves to partially undo the angling performed by the first wedge such that when focused, two spots form on the image plane. One spot is from the `on-axis' beam that passed straight through the hallowed portion of the wedges and another `off-axis' beam that passed through the both wedge prisms. Because the off-axis beam passed through the Alluxa filter at an angle, this beam is shifted toward the blue with respect to the on-axis beam. The `Y' shape is designed as such in order to create similar areas for the regions light passes through such that the on-axis and off-axis beams would be have roughly similar flux values\footnote{The original on-sky tests were to be performed using a telescope with a large secondary but were instead performed on one with a smaller secondary mirror. Since the prism proportions were originally estimated for a large secondary obstruction, this resulted in the on-axis beam's flux being larger than that of the off-axis beam.}. This took into account the fact that the image of the telescope pupil would be doughnut-shaped due to the obstruction of a secondary mirror.

The optical layout in panel (b) of Figure \ref{fig:wedge} was used in laboratory tests and on-sky tests. In order for the final image of the two beams to not overlap at the focal plane, the second wedge does not perfectly correct the angle induced by the first wedge. Originally, the wedge prisms were designed such that, if the Y-shape cuts were made on one wedge with its maximum thickness at 0 degrees, the complementary pair would have a matched shape when its maximum thickness is rotated by 15 degrees short of 180 degrees with respect to the 0 degree position of the first wedge. This 15 degree difference translates into a final angle between the two focusing beams of about 2 degrees. This angle was too large for the CCD camera used for our on-sky tests and so two wedges that were identical aside from a 180 degree rotation were used. To create the offset in angle, the second wedge was rotated 5 degrees with respect to the first wedge. This causes a slight misalignment in the Y-shapes. As a result, some light passing through the cut-out region of the first wedge passed through the Y-shaped glass of the second prism and vice-versa. Aside from a loss of photons, these secondary beams did not pose a problem because their wide 6 degree angle results in a focus far from the two primary beams we wish to image. This is also the case for the small amounts of light that pass through the edge of the first wedge and, due to the imposed angle in their direction and resulting change in vertical position, do not make it through the second wedge.

Although the wedge design is insensitive to a deviation in light due to the slight misalignment of the wedge prisms, changes in the position of the source image (i.e. a star) can move light from the `on-axis' beam to the `off-axis' beam. This could easily occur in on-sky tests as a result of guiding errors. A smaller source of this error can also be from changes in the distribution of light in the pupil image of telescope due to atmospheric turbulence. This scintillation noise would occur on fast time scales while we expect errors due to guiding to be either a slow variation if the telescope is allowed to drift, or, if a guider is used, varying on the timescales of guider corrections. We address these noise sources in on-sky tests and discuss solutions in the following sections.

\subsection{Laboratory Tests}
Before testing the oxyometer on sky, we performed tests in the laboratory in order to confirm the wavelength coverage of the instrument and to demonstrate good photometric performance. The laboratory setup was also useful to align all the optics. To create illumination at a similar focal ratio to that of the telescope we would use (f/6.3), we use a 100 mm lens to collimate the light from the halogen lamp and then a 150 mm plano-convex lens to create a converging beam. We use a super continuum laser as the light source to create bright spots that we record the spectrum of using our Ocean Optics spectrograph. The final spectra of the spots is very similar to what is shown in Figure \ref{fig:lab_setup}, however each bandpass FWHM is 0.03 nm wider and the off-axis beam is slightly redder, centered at 606.5 nm instead of the previous 605.75 nm. This is expected because the angle of incidence of the off-axis beam through the Alluxa filter reduced by 2 degrees compared to the initial bench-top setup. 

The super continuum laser was replaced by the halogen again as the light source and a series of 0.1~s exposure images were taken every 5~s for a duration of 45 minutes to test the photometric performance. At the noise levels observed, the halogen lamp appeared very stable and did not require any fits to remove the continuum variation. We use an SBIG ST-i camera for the imaging. With this data we found the standard deviation of $m_B - m_A$ to be 550 ppm, while the expected noise including photon, read, and dark noise was 575 ppm. The Allan deviation plot for these data shows a slope of -0.58 up to a bin size of 2.5 minutes, after which the standard deviation of the data binned by larger time intervals flattens out at about 100 ppm. This is likely due to variations in the halogen light source as there does appear to be some temporal structure in the data.  

\begin{figure*}
\centering
\begin{minipage}{.55\textwidth}
  \centering
  \includegraphics[width=0.99\linewidth]{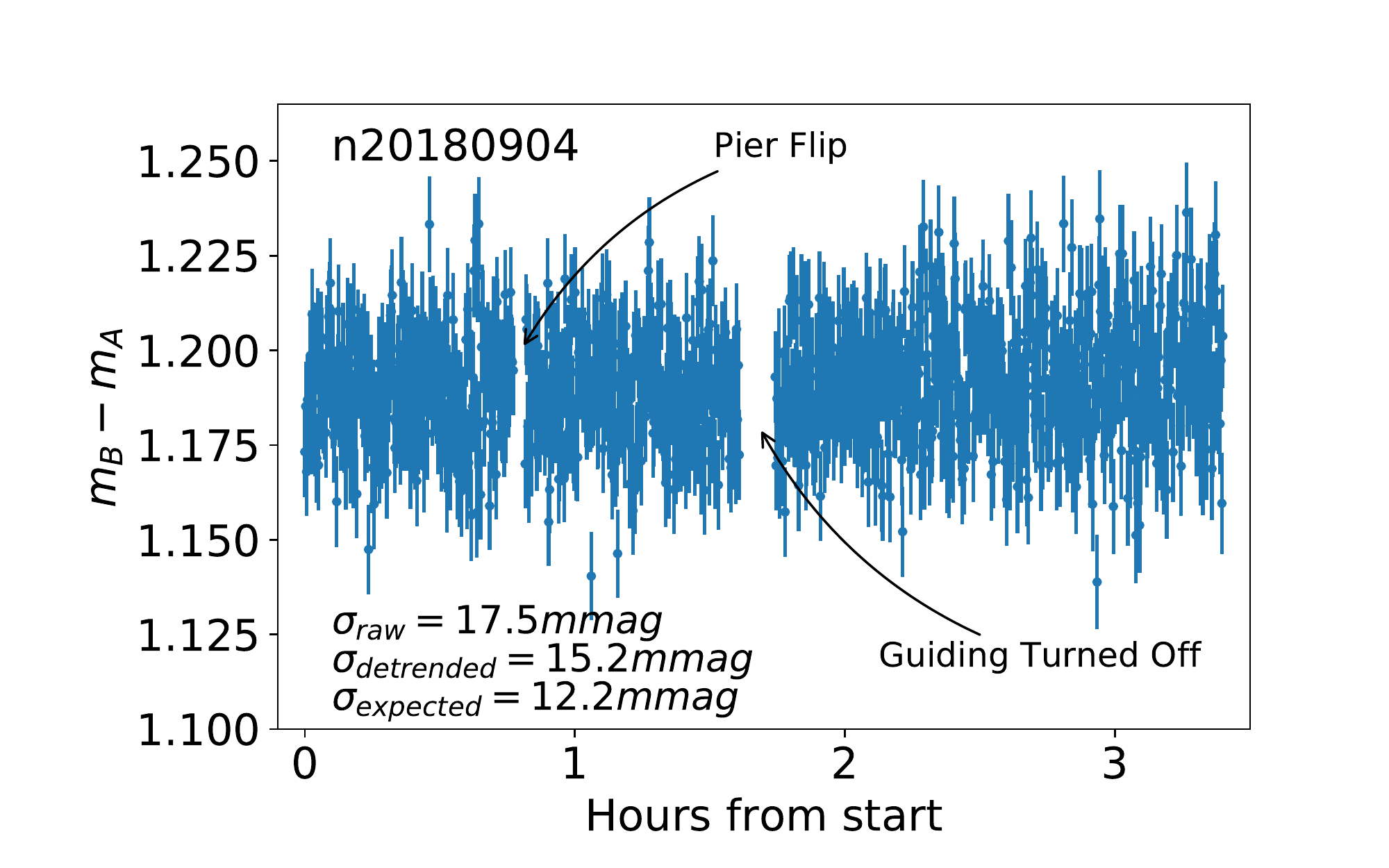}
    \caption{On-sky data run of Arcturus on the night of September 4th, 2018.}
    \label{fig:onskydata}
\end{minipage}
\hfill
\begin{minipage}{.43\textwidth}
  \centering
  \includegraphics[width=0.99\linewidth]{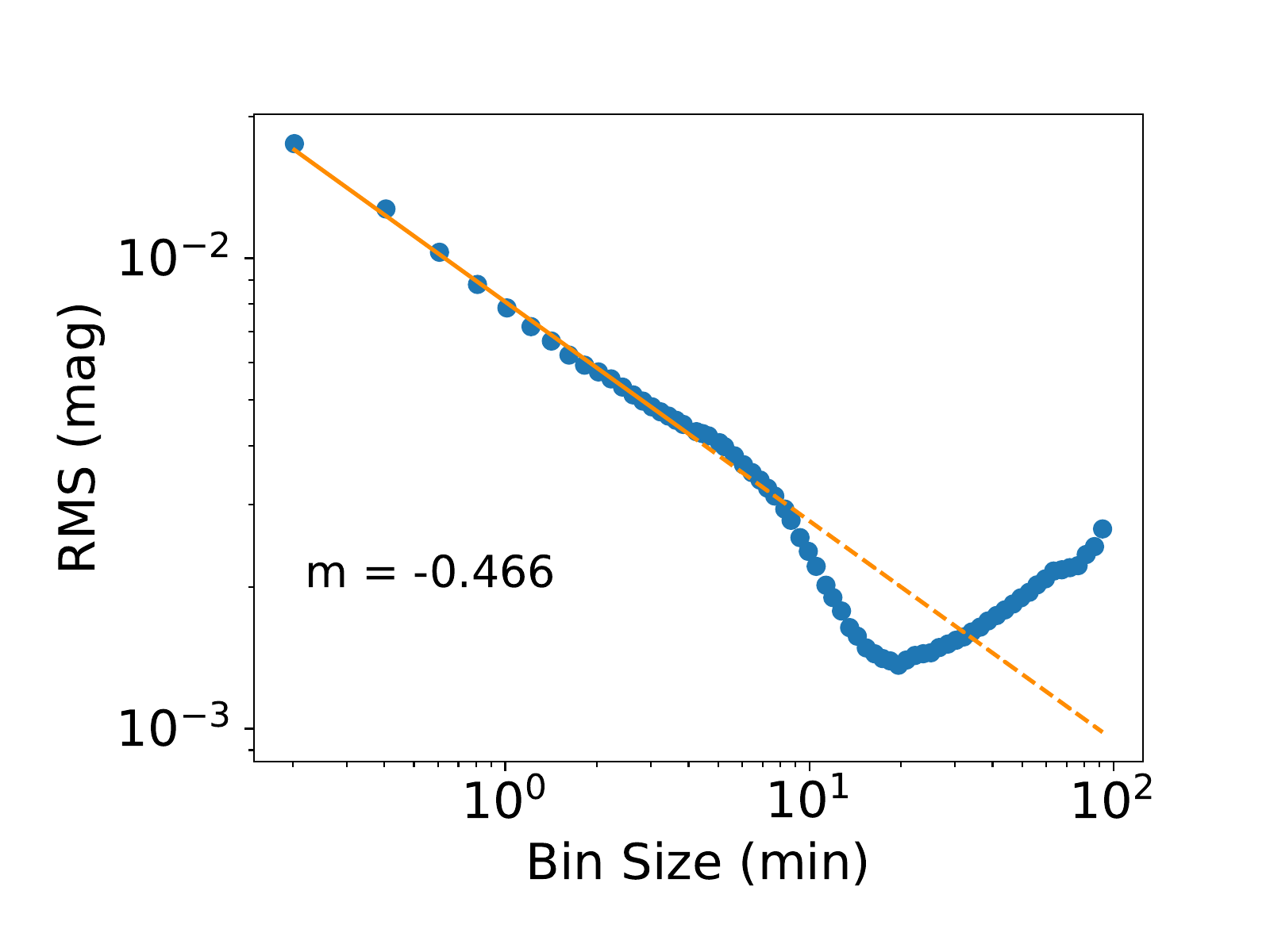}
    \caption{Allan deviation plot of the September 4th on-sky data run.}
    \label{fig:onskyallan}
  \end{minipage}
\end{figure*}

\subsection{On-Sky Tests}


To test the instrument on sky, we add a few more components to the simplified design presented in Figure \ref{fig:wedge}. We use a Paramount MyT mount with a 10" f/10 Meade Schmidt Cassegrain telescope. We attach to the back of the telescope a focal reducer to reduce the focal ratio to f/6.3, which sets the physical length of our instrument since we require our final beam diameter to be 1" to match the size of the optical components. After the focal reducer we attach a beam splitter to direct 8\% of the light to an SBIG ST-i camera that is used for guiding. The remaining 92\% of the light passes through the collimating lens (focal length 150 mm), then the first wedge prism followed by the Alluxa filter, and then the second wedge prism. The light is then focused with a 60 mm lens onto an SBIG 8300M camera that is cooled to 0 degrees Celcius. An image of the on-sky setup mounted on the Meade 10" telescope is shown in Figure \ref{fig:onsky}. The focus onto the final camera is adjustable, but we keep the final image out of focus, which results in two images resembling the wedge prism. An example final image with the corresponding photometric apertures is shown in Figure \ref{fig:onskyspot}, where we have labeled which image is the on-axis, relatively redder bandpass and which is the off-axis, bluer bandpass. 

Using the wedge prism and Alluxa filter setup on the 10" Meade Telescope at the University of Pennsylvania, we imaged Vega on September 4th, 2018 in continuous 10s exposures over a 3.5 hour period. We used Software Bisque's program TheSkyX for telescope control, CCD control, and guiding. Dark frames were taken before the data run at the same exposure time and CCD operating temperature and used to create a median frame that was subtracted off each science frame. The fluxes were extracted and the sky-background estimated from a concentric annulus. The sky-background levels differed across the CCD but were small (5-25 counts in the total aperture area, increasing with airmass) and therefore did not significantly affect our measured flux ratio. The ratio of the final dark and sky-subtracted fluxes for each band for the 3.5 hour of measurements are shown in Figure \ref{fig:onskydata}. The corresponding Allan deviation plot of these data is shown in Figure \ref{fig:onskyallan}. The RMS of the binned data decreases like white noise (slope of -0.47) up to a bin size of 10 min, dropping to a minimum RMS of 1 mmag.

The expected noise due to CCD noise and photon noise is calculated in the same way as for the laboratory tests, except we include noise contributions due to atmospheric scintillation. The standard deviation of the Vega observations is 17.5 mmag, which is 1.4 times the expected noise of 12.2 mmag. While this is a small discrepancy, we observe that the flux ratio correlates with the background sky counts at a fixed airmass. This is unexpected and we discuss possible explanations below. We detrend the flux ratio using the ratio of the background sky counts and find that the scatter in the data reduces to 15.2 mmag, or 1.2 times the expected noise. At around 45 min into the observing run a pier flip occurred requiring that the star be recentered. This caused a slight shift in the mean of the data before and after we performed the pier flip. To correct this, we shift the first $\sim$30 min of the data so the median values match in the two regimes. Additionally, at around 1.6 hours into the observing run we turn off the guiding to see how the flux ratio and the scatter in the data are affected by a small drift in the star. We find that the flux ratio remains flat although the noise increases about 90 minutes after the guiding was turned off. As the illumination of the wedge prisms slowly changes as the star gradually drifts, we might expect the flux ratio to also drift as light falling on the edge of the wedge prisms moves between bands. This could be occurring at a low level, or be directly causing the increased scatter observed after guiding is turned off. 


We consider a few explanations for the correlation between the flux ratio and the sky background flux ratio. First, we observe that the two quantities are anti-correlated and that the correlations are present with or without guiding. Therefore, the possibility of a changing CCD sensitivity or tracking errors are both ruled out as potential causes. The possibility of a low-level ghost image is also unlikely since none was observed in the co-added frame of all the exposures or in the lab where a brighter light source was used. Because the effect changes strengths with varying airmass, the correlation is likely to be related to the changing illumination of the instrument and varying background sky brightness and/or the position of the moon. This could be confirmed with further tests.

Overall, on-sky tests showed that an instrument with our oxyometer design can be used to observe an object simultaneously in two ultra-narrow bands near the photon limit. Observing with the instrument is fairly simple, although splitting the telescope pupil necessitates stable instrument illumination. With the guiding capabilities of our off-the-shelf My-T mount, we were able to produce a stable illumination to reach near photon limit, but precise guiding and an AO system could further help improve the precision of the measurements and reduce the effects of scintillation noise.

\section{Discussion}\label{sec:discussion}

\subsection{Alternative Design Directions}
\begin{figure}
    \centering
	\includegraphics[width=0.6\linewidth]{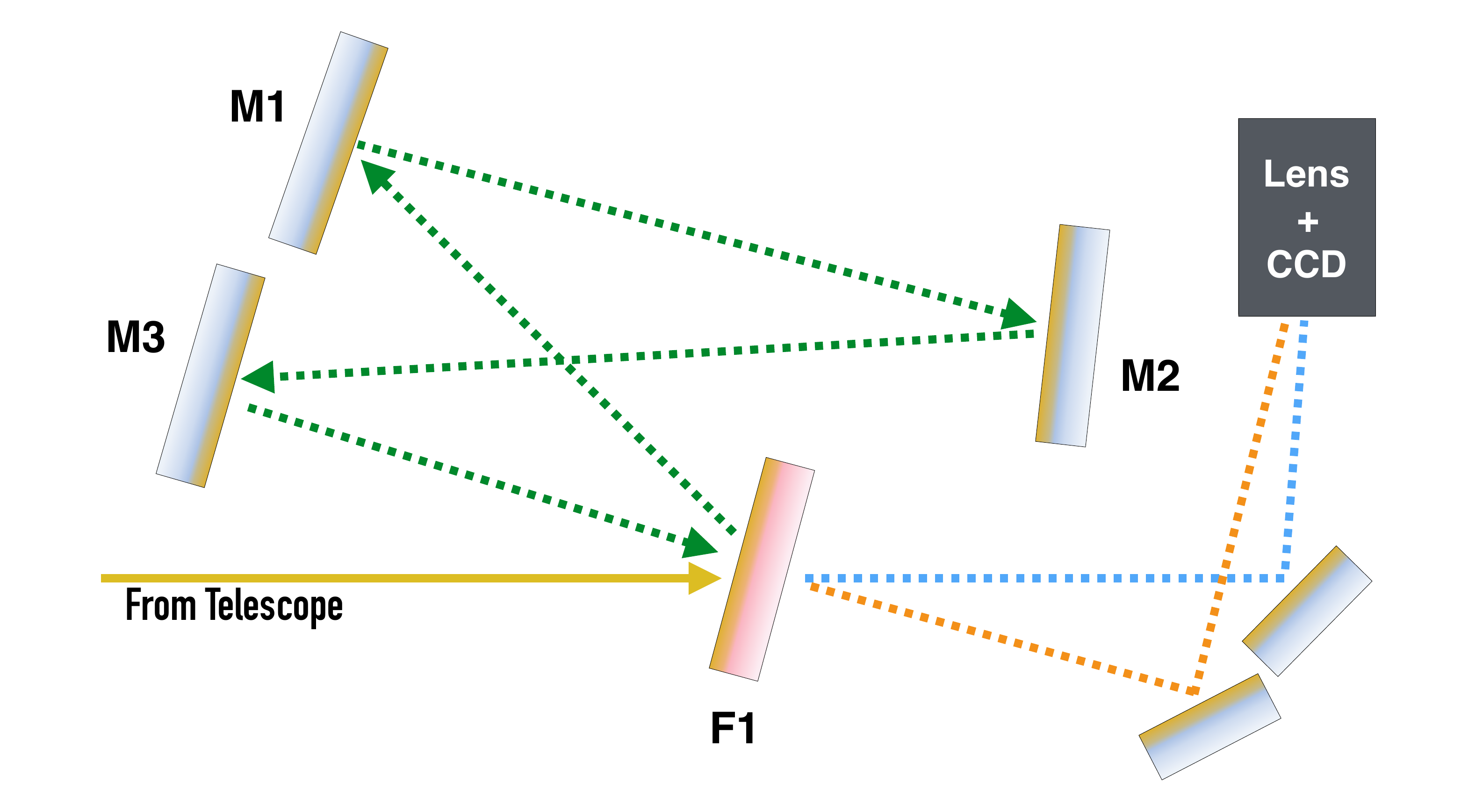}
    \caption{Alternate design for an oxyometer relying that avoids a 50\% cut in photons due to splitting the collimating as in the previous designs considered. Collimated light from the telescope passes through a tilted narrow-band interference filter experiencing a blueshifted effective bandpass. The light that is rejected is reflected off several mirrors (M1-M3) back through the filter at normal incidence so that this second beam redder than the first. Both beams can then be directed with more mirrors towards a lens and camera system for imaging.}
    \label{fig:all_reflective}
\end{figure}

As was shown in \S \ref{sec:noise}, maximizing the overall throughput of our oxyometer is essential to pushing down the photon noise floor. Currently the designs tested split the telescope pupil in half to perform measurements in the two different wavelength bands simultaneously. While these designs were simple and allowed for an easy on-sky demonstration, an instrument that avoids the 50\% beam split is certainly preferred. We present a design that achieves this and the goal of simultaneous imaging in Figure \ref{fig:all_reflective}. In this setup, a collimated beam from the telescope passes through the filter at an AOI of 6-8 deg, reflects off the filter and then, with a series of mirrors, the rejected initial reflection is redirected back through the filter at normal incidence. Mirrors can then direct the two outgoing beams, now at the desired wavelengths, to a lens and CCD system.

Although it is possible to clock the first mirror, M1, in the design such that the reflecting beam passes immediately back through the filter at an AOI different from the initial beam by 6 to 8 deg, this would require two AOIs with respect to the filter both greater than zero to fit the reflecting beam through the small filter diameter. In this case, a filter optimized for high throughput at an AOI away from normal incidence would be required to avoid beam degradation. We prefer a design where the two mirrors, M2 and M3, are added because this allows for more flexibility in the spacing of the optics and, with high reflectivity mirrors, will not sacrifice instrument throughput. Additionally, mirrors M2 and M3 could even be replaced with more ultra-narrowband filters for more spectral coverage. With custom designed holders for the filter and mirrors, it would also be possible to connect this instrument directly to a telescope, similar to what was done in \S \ref{sec:onsky}. This will remove optical fiber dependencies which will lower the throughput of the instrument and can impose chromatic modal noise.

This design also does not require a minimum beam diameter, which helps keep the design compact. It additionally benefits from producing the same pupil image for each bandpass meaning scintillation noise will be correlated between photometric bands and not add to the noise of the flux ratio. As before, the largest impact on overall throughput are the CCD QE and the telescope reflectivity. Current EMCCDs can reach \textgreater 95\% QE and telescope reflectivity can reach over 95\% efficiency at red to NIR wavelengths \citep{gemini_mirror}. With these specifications, reaching the 80\% efficiency assumed in the right panel of Figure \ref{fig:hours_snr3} is feasible.

\subsection{Observing Other Absorbing Species}
The design of our oxyometer relies upon a narrow-band interference filter centered around the 760 nm oxygen bandhead. Since these filters can be manufactured with a wide range of central wavelengths from the UV to NIR, the general oxyometer design can be easily tuned to target other absorbing species by replacing the filter with one manufactured to overlap the molecular transition of interest. Alluxa is one of several companies that manufactures filters ranging from 300 nm to 1.6 microns central wavelength and can custom manufacture a filter with a specified bandpass. Other molecular transitions in both hot Jupiters and terrestrial planets could be assessed to determine their suitability for observations with an oxyometer-style instrument. With their larger scale heights, hot Jupiter exoplanets are easier targets and a high throughput spectrophotometer would allow smaller telescopes to contribute to atmospheric characterization surveys of easier targets. More sampling of transits would not only improve our measurement accuracy but would also be useful for tracing signatures of stellar contamination that could evolve between transits. With a simple instrument like an
oxyometer adapted to characterizing species present in hot Jupiters, current facilities down to apertures as small as a few meters in diameter could significantly contribute to surveys of hot Jupiter atmospheres.



\section{Conclusions} \label{sec:conclude}
We present the motivation and design for an ultra-narrowband photometer, which we call an oxyometer, to perform simultaneous observations of a star around the oxygen 760 nm bandhead for the detection of oxygen in terrestrial exoplanet atmospheres. We showed that spectrophotometry with such an instrument installed on a large aperture telescope with two bandpasses as narrow as 0.3 nm FWHM could be used for making a ground-based detection of exoplanetary oxygen around nearby M dwarf stars that have a radial velocity greater than 75~km/s towards Earth such that the telluric oxygen absorption does not overlap the oxygen bandhead in the exoplanet absorption spectrum. In space no telluric oxygen will need to be avoided and in this case the full 2 nm width oxygen bandhead can be probed making a 10 m telescope capable of probing M4 dwarfs or cooler that are nearer than 5 pc. We show that increasing the throughput of the instrument can significantly reduce the required number of transits to reach a 3$\sigma$ detection. 

To demonstrate the ability to build an instrument that maximizes throughput while achieving the assumed spectral coverage required for detecting the oxygen signal, we used a narrowband interference filter of 0.3 nm FWHM. With two different optical designs we were able to split the light source and send half of the light through the interference filter at normal incidence and the other half through at a 6 to 8 degree AOI. The first design was constructed on an optical bench and was very large so we designed custom wedge prisms that compactified the instrument and allowed it to be attached to a telescope and tested on the sky. With the on-sky instrument we demonstrated the ability to achieve simultaneous photon-limited photometry in two ultra-narrow bands spaced 1~nm apart. Since interference filters of the type used here can be ordered in custom wavelengths and bandpasses, instruments of a similar design to the oxyometer could be constructed to measure a different molecular transition by shifting the central wavelength of the filter. Future work on this instrument will explore an all-reflective design that will increase the instrument throughput by a factor of two and is flexible to include more spectral coverage. A multi-band photometer for transit spectrophotometry that achieves high spectral resolution (R$\sim$2500) in addition to high throughput would reduce the required observing time for the most challenging targets and make easier measurements accessible to smaller aperture telescopes.

The ability to characterize terrestrial planets in order to assess their habitability is a major, long-term goal of the field. With upcoming missions such as TESS that are expected to find many transiting planets around cooler stars than the sun, we will soon have large samples of terrestrial exoplanets for which atmospheric characterization will be crucial for achieving this goal. Building highly specialized instruments for characterizing the atmospheres of the these targets can complement current methods by focusing exclusively on specific atmospheric species such as molecular oxygen.

\section*{Acknowledgements}
The authors would like to thank the referee for his or her thorough comments that greatly improved this manuscript. We thank David Sliski for his extensive work in setting up the MyT mount in the Penn observatory and also for helping me in setting up for on-sky tests on multiple occasions. We also thank Nikku Madhusudhan, Luis Welbanks, and Jasmina Blecic for helping us with generating an Earth atmosphere transmission spectrum. Thanks to Andy Szentgyorgyi for insightful comments on the oxyometer concept. We also thank Peter Harnish and Bill Berner for loaning us various gadgets, including the Ocean Optics spectrograph that was used frequently for this work. We thank Harold Borders for taking the time to give us advice and materials for stabilizing the oxyometer on-sky setup and Ben Rackham and Daniel Apai for providing us with their high resolution models of stellar contamination of transit spectra to integrate over our unique bandpasses.

This material is based upon work by ADB supported by the National Science Foundation Graduate Research Fellowship under Grant No. DGE-1321851. This work was also performed, in part, by SPH under contract with the Jet Propulsion Laboratory (JPL) funded by NASA through the Sagan Fellowship Program executed by the NASA Exoplanet Science Institute.

\bibliography{oxyometer}   
\bibliographystyle{aasjournal}   
\end{document}